\documentclass[11pt,a4paper]{article}
\usepackage{jheparxiv}
\usepackage{amsmath}

\usepackage{amsfonts}
\usepackage{amssymb}
\usepackage{latexsym}
\usepackage{mathrsfs}
\usepackage{color}
\usepackage{cancel}
\usepackage{xcolor}
\usepackage{slashed}
\usepackage{tikz-cd}
\usepackage{mathtools}
\usepackage{bm, bbm}
\usepackage{float}
\usepackage[]{todonotes}
\usepackage{tikz-feynhand}
\usepackage{physics}
\usepackage{booktabs}
\usepackage{comment}
\usepackage{tcolorbox}
\allowdisplaybreaks[4]

\makeatletter
\renewcommand{\todo}[2][]{%
    \@todo[caption={#2}, #1]{\begin{spacing}{0.5}#2\end{spacing}}%
} 
\makeatother 
\usepackage{setspace}
\renewcommand{\[}{\begin{equation}\begin{aligned}}
\renewcommand{\]}{\end{aligned}\end{equation}}
\newcommand{\nn}{\nonumber \\}

\newcommand{\Amp}{ \mathcal{A} }

\newcommand{\hd}{\hat{\rm d}}
\newcommand{\hdelta}{\hat{\delta}}

\newcommand{\ima}{{\rm Im}\,}

\title{\scalebox{0.93}{Quantum Effects for Black Holes with On-Shell Amplitudes}}
\author[a]{\scalebox{0.98}{Katsuki Aoki,}}
\affiliation[a]{Center for Gravitational Physics and Quantum Information,
Yukawa Institute for Theoretical Physics, Kyoto University, 606-8502, Kyoto, Japan }
\emailAdd{katsuki.aoki@yukawa.kyoto-u.ac.jp}
\author[a]{\scalebox{0.98}{Andrea  Cristofoli,}} 
\emailAdd{cristofoli@yukawa.kyoto-u.ac.jp}
\author[b,c,d]{\scalebox{0.98}{Hyun Jeong,}}
\affiliation[b]{Kavli IPMU (WPI), UTIAS, The University of Tokyo, Kashiwa, Chiba 277-8583, Japan}
\affiliation[c]{\textit{Research Center for the Early Universe (RESCEU), Graduate School of Science,  The University of Tokyo, Tokyo 113-0033, Japan}}
\affiliation[d]{\textit{Department of Physics, Graduate School of Science, The University of Tokyo, Tokyo 113-0033}}
\emailAdd{jeong\_hyun@resceu.s.u-tokyo.ac.jp}
\author[e]{\scalebox{0.98}{Matteo Sergola,}}
\affiliation[e]{Institut de Physique Theorique, CEA, CNRS,
Universite Paris-Saclay, F–91191 Gif-sur-Yvette cedex, France}
\emailAdd{matteo.sergola@ipht.fr}
\author[f]{\scalebox{0.98}{Kaho Yoshimura}}
\affiliation[f]{Graduate School of Arts and Sciences, University of Tokyo, Komaba, Meguro-ku, Tokyo 153-8902}
\emailAdd{yoshimura-kaho848@g.ecc.u-tokyo.ac.jp}

\abstract{We develop a framework based on modern amplitude techniques to analyze emission and absorption effects in black hole physics, including Hawking radiation. We first discuss quantum field theory on a Schwarzschild background in the Boulware and the Unruh vacua, and introduce the corresponding $S$-matrices. We use this information to determine on-shell absorptive amplitudes describing processes where a black hole transitions to a different mass state by absorbing or emitting quanta, to all orders in gravitational coupling. This on-shell approach allows for a universal description of black holes, with their intrinsic differences encapsulated in the discontinuities of the amplitudes, without suffering from off-shell ambiguities such as gauge freedom. Furthermore, the absorptive amplitudes serve as building blocks to describe physics beyond that of isolated black holes. As applications, we find that the Hawking thermal spectrum is well understood by three-point processes. We also consider a binary system and compute the mass shift of a black hole induced by the motion of a companion object, including quantum effects. We show that the mean value of the mass shift is classical and vacuum-independent, while its variance differs depending on the vacuum choice. Our results provide confirmation of the validity of the on-shell program in advancing our understanding of black hole physics.

}

\begin{document}
{\baselineskip0pt
\rightline{\baselineskip16pt\rm\vbox {
           \hbox{YITP-25-143}
           \hbox{UT-Komaba/25-9}
           \hbox{IPMU25-0045}
           \hbox{RESCEU-19/25}
}}%
}
\maketitle

\section{Introduction}

Chandrasekhar famously remarked that \emph{``the black holes of nature are the most perfect macroscopic objects there are in the universe: the only elements in their construction are our concepts of space and time''} \cite{Chandrasekhar:1985kt}. Conversely, on-shell scattering amplitudes have been described as \emph{``the most perfect microscopic structures in the universe''} \cite{Dixon:2011xs}, grounded in relativity, unitarity, and causality. In recent years, these two extremes of perfection have intriguingly come together.

Applied to the gravitational two-body problem, quantum scattering amplitudes have shown that, after taking the classical limit, it is possible to perform novel calculations in general relativity for classical black hole dynamics, as first demonstrated in \cite{Bern:2019crd, Bern:2019nnu}.
At the heart of this union lies the reformulation of QFT, often referred to as \emph{the on-shell program}. In this modern approach to QFT, no place is given to Feynman diagrams or traditional perturbation theory; instead, the approach relies entirely on modern tools from particle physics, such as generalized unitarity \cite{Bern:1994zx,Bern:1994cg, Bern:1995db,Bern:1997sc, Britto:2004nc}, the double copy \cite{Kawai:1985xq, Bern:2008qj,Monteiro:2014cda}, and integration-by-parts \cite{Laporta:2000dsw, Chetyrkin:1981qh}. Computing classical gravity and black hole (BH) observables using these techniques is now a well-established field, demonstrating how fruitful and impressive the cooperation between different research areas can be. For reviews of applications to classical gravitational-wave physics, see \cite{Kosower:2022yvp, Buonanno:2022pgc,Bjerrum-Bohr:2022blt}; some recent developments can be found in~\cite{Bini:2023fiz,Caron-Huot:2023ikn,Bautista:2023sdf,Georgoudis:2023eke,Luna:2023uwd,Cangemi:2023bpe,Bini:2024rsy,Aoki:2024bpj,Dlapa:2024cje,Brunello:2024ibk,Damgaard:2024fqj,Gambino:2024uge,Kim:2024grz,Alaverdian:2024spu,Buonanno:2024byg,Bern:2024adl,Cheung:2023lnj,Kosmopoulos:2023bwc,Akpinar:2024meg,Akpinar:2025bkt,Falkowski:2024bgb,Bautista:2024agp,Bern:2024vqs,Solon:2024zhr,Elkhidir:2024izo,Iteanu:2024dvx,Bohnenblust:2024hkw,Driesse:2024feo,Kim:2025hpn,Heissenberg:2025ocy,Guevara:2024edh,Cristofoli:2025esy,Caron-Huot:2025tlq,Vazquez-Holm:2025ztz,Ivanov:2025ozg,Klisch:2025mxn,Bini:2025vuk,Brammer:2025rqo,Akhtar:2025nmt,Alessio:2025nzd,Georgoudis:2025vkk,Dlapa:2025biy,Aoude:2025xxq}.

Nevertheless, scattering amplitudes are quantum at heart. This simple observation indicates that there should be another side to the program of applying modern tools from QFT to \emph{classical} gravitational physics. The fact that black holes evaporate by \emph{quantum} effects was first discovered by Hawking in a pioneering series of papers \cite{Hawking:1974rv,Hawking:1975vcx}, paving the way for entirely new research fields to develop (see, for instance, the review \cite{Page:2004xp}). These original observations took place more than 50 years ago, when very little was known about the use of amplitude techniques to gravitational physics~\cite{Iwasaki:1971vb}. Equipped with the modern on-shell techniques and knowledge learned from their applications to classical gravitational physics, it is natural to ask the following: \emph{in the quantum realm, can on-shell amplitudes advance our understanding of black holes?}

The first task in addressing this question would be to understand how to deal with the classical horizon effects from a scattering perspective. At the classical level, the horizon is an absorbing boundary and the black hole increases in its mass by absorption. This process can be understood as amplitudes with external particles of different masses and spins responsible for the initial and final black holes~\cite{Aoude:2023fdm, Jones:2023ugm, Chen:2023qzo, Aoki:2024boe, Aoude:2024jxd, Bautista:2024emt} (see also~\cite{Goldberger:2005cd, Porto:2007qi} for the worldline approach where the horizon effects are described by interactions with ``invisible'' gapless modes localized on the worldline). These amplitudes can be used to describe horizon absorption effects in binary black hole dynamics. The absorption effects in observables are highly suppressed when binaries are far separated, but they are nonetheless present and possess a classically relevant contribution. As for the quantum nature of black holes, Hawking effects have been investigated based on the worldline~\cite{Goldberger:2020wbx,Goldberger:2020geb,Kim:2020dif,Ilderton:2025umd}, while fewer developments exist from the amplitude perspective. For instance, black hole formation processes in super-Planckian scattering are investigated in~\cite{Giddings:2007qq, Giddings:2009gj}. In \cite{Aoude:2024sve}, the frameworks of the eikonal amplitude exponentiation and the KMOC formalism~\cite{Kosower:2018adc} were used to derive the thermal distribution and temperature of a Schwarzschild black hole.

In this paper, we aim to strengthen the direction of the on-shell approach to black holes by bridging the gap between  amplitudes and traditional approaches to Hawking radiation and QFT on curved spacetime. The $S$-matrices in the black hole \emph{spacetime} can be defined through the Bogoliubov transformation~\cite{dewitt1975quantum, Frolov:1998wf}. These $S$-matrices are matched to the on-shell scattering amplitudes of black hole \emph{particles} defined in the boundary Minkowski spacetime. This allows us to incorporate quantum effects of black holes into the modern amplitude program, in a similar way to the on-shell description of the classical horizon~\cite{Aoude:2023fdm, Jones:2023ugm, Chen:2023qzo, Aoki:2024boe, Aoude:2024jxd, Bautista:2024emt}. In this picture, for instance, the Hawking radiation can be seen as a decay process of a black hole into a smaller-mass black hole, rather than the conventional picture of vacuum particle production. Combined with the amplitude techniques, the constructed amplitudes can be employed to study more complex dynamics, such as binary dynamics, providing a new avenue to explore quantum effects of black holes.

Our work is organized as follows. In Section~\ref{sec:Quantum_fields_in_Schwarzschild}, we study the \(S\)-matrix in a black-hole background and analyze two physically distinct vacua --- Boulware and Unruh --- as our particular interest. To connect this analysis with modern on-shell techniques, in Section~\ref{sec:on-shell}, we build scattering amplitudes that treat black holes as on-shell one-particle states with a continuous spectrum. Horizon absorption and Hawking radiation are encoded in on-shell three-point amplitudes whose structures are fixed by the symmetry of the system and by matching to the black hole \(S\)-matrix. Within this framework, the black holes are considered as composite one-particle states, and the ``invisible sector'' encoded in the black hole \(S\)-matrix emerges naturally as their internal degrees of freedom. In Section~\ref{sec:observables}, we then discuss observables, the Hawking thermal spectrum and quantum effects on a binary system, based on our on-shell approach. We conclude in Section~\ref{sec:conclusions}. In Appendices~\ref{sec:Thermal_Phenomenon_as_a_squeezed state} and~\ref{sec:Bogoliubov}, we outline thermodynamics with squeezing, Bogoliubov transformations, and the construction of the $S$-matrix.

\subsection*{Conventions}{ We work in four dimensions with the mostly negative signature $(+,-,-,-)$. Factors of $2\pi$ are absorbed as in \cite{Kosower:2018adc} $\hat{\dd}^np := \dd^n p /(2\pi)^n$ and $\hat{\delta}^n(\cdot) := (2\pi)^n \delta^n(\cdot)$.  We also define phase space volumes as 
\[
\dd\Phi(p):=\hat{\dd}^D p \,\hat{\delta}^D(p^2-m^2)\Theta(p^0), \,\,\,\, \dd\Phi(p_1, \cdots,  p_n):=\prod_{i=1}^n \dd\Phi(p_i).
\]
 Unless specified, limits of the integral are from $-\infty$ to $+\infty$. We set $c=1$, keep $G$ explicit and reintroduce $\hbar$ whenever relevant. While multiple labels are often required to specify quantities (e.g.,~creation/annihilation operators for different modes), we sometimes omit labels irrelevant to discussions unless confusion arises.

\section{Quantum fields on Schwarzschild spacetime}\label{sec:Quantum_fields_in_Schwarzschild}

Physical processes in Schwarzschild spacetime are twofold. First, there is the scattering of waves off the gravitational potential created by the black hole and the absorption by the event horizon. They occur even in the scattering of classical waves. The second is a thermal or quantum effect induced by the horizon, namely, Hawking radiation. The latter is understood as a change of vacuum through a Bogoliubov transformation. The Bogoliubov transformation is expressed by a squeezing operator, which can be further identified with the $S$-matrix of a quantum field in the Schwarzschild background~\cite{dewitt1975quantum, Frolov:1998wf}. This $S$-matrix has the complete information on scattering, absorption, and emission by the black hole within the black hole perturbation theory (BHPT). Indeed, we will see in Sec.~\ref{subsec:Hawking_radiation_as_a_resummation_of_amplitudes} that the Hawking thermal distribution arises from summing over transition amplitudes of particle creation. Note that this description of a thermal system is known as thermo-field dynamics~\cite{Umezawa:1982nv,Umezawa:1993yq,gerry2023introductory} where the Hilbert space is enlarged and the thermal effect is described by the entanglement of states with pair production of particles. See Appendix~\ref{sec:Thermal_Phenomenon_as_a_squeezed state} for a brief review. In this perspective, the enlarged Hilbert space is understood as a direct product of the spaces of particles at infinity and at the horizons~\cite{Wald:1984rg, Frolov:1998wf}.

\subsection{Scalar modes on Schwarzschild spacetime}
\label{sec:mode}
We start by reviewing the mode functions of a quantum field in the Schwarzschild spacetime. We mainly follow Chapter 10 of~\cite{Frolov:1998wf} and refer the readers there for further details. We first focus on a massless scalar field and revisit the graviton case in Sec.~\ref{subsec:Graviton_modes_on_Schwarzschild spacetime}.

The analysis is based on the semi-classical approach, where the black hole is treated as a fixed classical background. The Schwarzschild metric is given by
\begin{align}\label{eq:Schwarzschild_metric}
    \dd s^2 &=\left(1-\frac{2GM}{r}\right)\dd t^2 - \left(1-\frac{2GM}{r}\right)^{-1} \dd r^2 - r^2 \dd \Omega^2
\nn
&= \left(1-\frac{2GM}{r}\right) \dd v^2 - 2 \dd v \dd r - r^2 \dd \Omega^2
\,, 
\end{align}
where $u=t-r_*$ and $v=t+r_*$ with $r_*=r+2GM \ln |r/(2GM)-1|$ denote the advanced and retarded time, while $M$ is the mass of the black hole. The line element in \eqref{eq:Schwarzschild_metric}, is in Schwarzschild coordinates, which only covers the exterior of the black hole (region I), while the second line, the ingoing Eddington-Finkelstein coordinates, covers both inside and outside of the black hole (regions I and II). See the left panel of Fig.~\ref{fig:penrose_diagram}. Here we will not discuss the maximal extension because only regions I and II are relevant to a black hole formed by a gravitational collapse.
\begin{figure}[t]
    \centering
    \includegraphics[width=\linewidth]{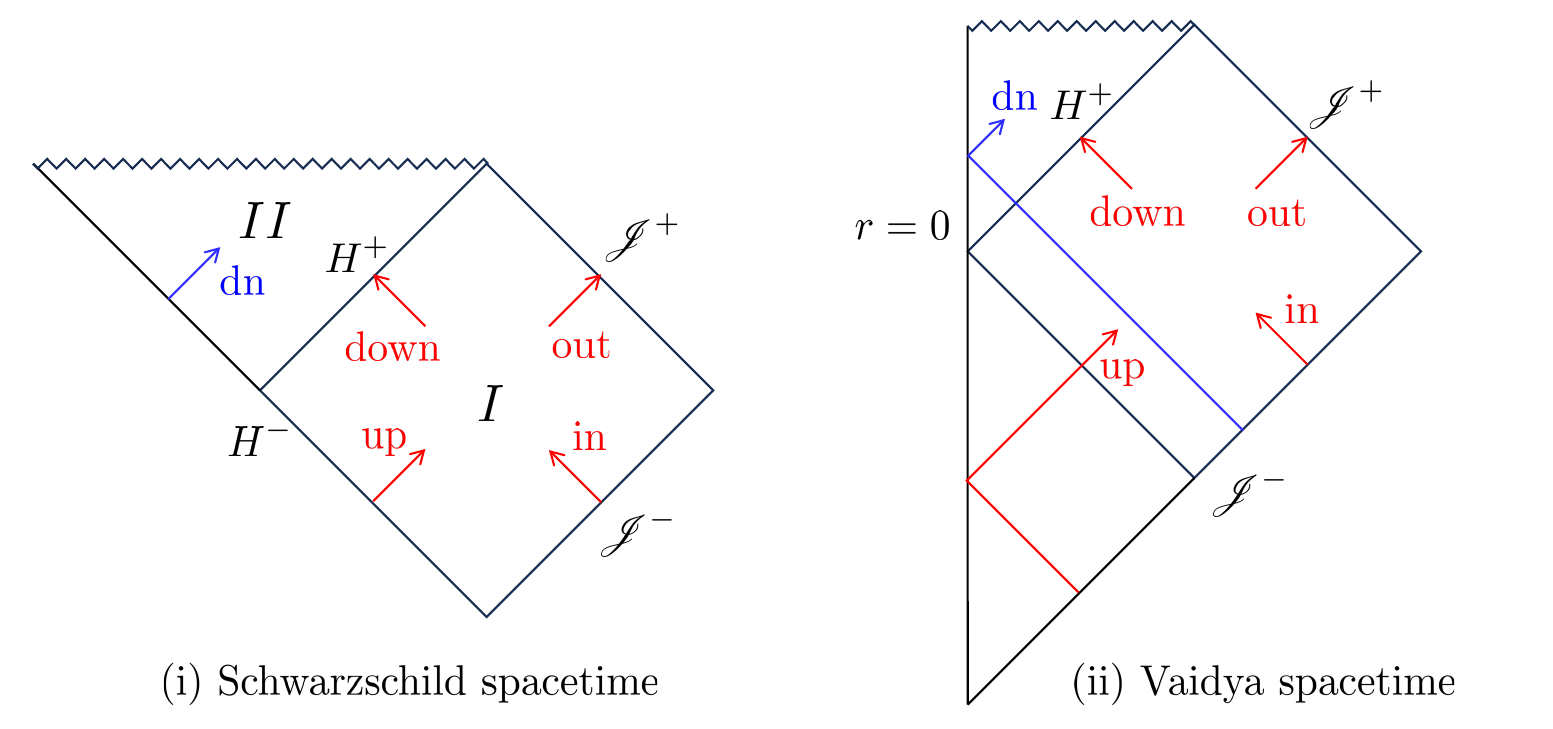}
    \caption{Carter-Penrose conformal diagram for Schwarzschild (in the ingoing Eddington-Finkelstein coordinates) and Vaidya spacetimes. The arrows represent the boundary conditions of: in, out, up, down, and dn modes. The Boulware vacuum can be constructed only by the mode outside of the horizon (red), while the Unruh vacuum, which agrees with the vacuum of the Vaidya spacetime, requires the ``dn" mode (blue).}
    \label{fig:penrose_diagram}
\end{figure} 
The massless scalar field obeys the Klein-Gordon equation
\begin{equation}
    \Box \varphi = 0. 
\end{equation}
Exploiting the static and spherically symmetric nature of the background, we can separate the variables as
\begin{equation}
    \varphi_{J} = e^{-i\omega t}\frac{u_l (r)}{r} Y_{lm} (\theta, \phi)\,.
\end{equation}
Here $J$ denotes the collective index $\{\omega lm\}$, and $Y_{lm} (\theta, \phi)$ is the spherical harmonics. 
The function $u_{l} (r)$ obeys the following radial equation
\begin{equation}
\label{eq:radial_equation}
    \left[\frac{d^2}{dr_*^2} + \omega^2 - V_l (r)\right] u_l(r) = 0\,, 
\end{equation}
where the potential is 
\begin{equation}
    V_l(r) = \left(1-\frac{2GM}{r}\right) \left(\frac{l(l+1)}{r^2} + \frac{2GM}{r^3} \right)\,.
\end{equation}

The general solution to the radial equation~\eqref{eq:radial_equation} in the exterior region of a black hole (region $I$) can be expressed as a linear combination of any two of the following mode functions: in, up, out, and down. As illustrated in Fig.~\ref{fig:penrose_diagram}, the four key mode functions are each picked out by a simple boundary condition\footnote{Note that while we are interested in boundary conditions at $\mathscr{J}^\pm$, what we usually work with is the large $r$ limit of the radial part of the wave equation in $(t,r,\Omega)$ coordinates. This is a standard approach in dealing with radiation processes. Implicitly, since we are looking at boundaries on the Penrose diagram, this also requires that $t$ is such that $t-r_*$ or $t+r_*$ is finite.}:

\begin{itemize}
  \item The \textbf{``in''} mode, $\varphi_{{\rm in},J}$, is purely ingoing at past null infinity, $\mathscr{J}^-$.
  \item The \textbf{``up''} mode, $\varphi_{{\rm up},J}$, is purely outgoing at the past horizon, $H^-$.
  \item The \textbf{``out''} mode, $\varphi_{{\rm out},J}$, is purely outgoing at future null infinity, $\mathscr{J}^+$.
  \item The \textbf{``down''} mode, $\varphi_{{\rm down},J}$, is purely ingoing at the future horizon, $H^+$.
\end{itemize}
Each mode function is tailored to satisfy exactly one of these ingoing or outgoing conditions, making them the natural bases for scattering and horizon analyses. In particular, such solutions have the following inner product
\begin{align}
(\varphi_1, \varphi_2):=i \int_{\Sigma} \dd \Sigma^{\mu}(\bar{\varphi}_1 \partial_{\mu} \varphi_2 - \varphi_2 \partial_{\mu} \bar{\varphi}_1 )
\,,
\end{align}
where the bar denotes complex conjugation and $\dd \Sigma^{\mu}$ is the future-directed vector of the Cauchy surface $\Sigma$, leading to the normalization for same boundaries
\begin{align}
(\varphi_J,\varphi_{J'})=\hdelta_{J,J'}:=2\omega \hdelta(\omega-\omega')\delta_{ll'}\delta_{mm'}
\,. 
\label{normalisation}
\end{align}
Note that the Fourier mode has the infinite norm due to the delta function in \eqref{normalisation} and the precise treatment requires the wave-packet states integrated over a certain frequency interval or confining the system to a box in a finite time interval. We will revisit this issue when computing the thermal spectrum. 
Since the ``in" and ``up" modes are purely ingoing and outgoing in the past Cauchy surface, they are orthogonal; similarly, the ``out" and ``down" modes are orthogonal:
\begin{align}
(\varphi_{{\rm in}, J},\varphi_{{\rm up}, J'})=(\varphi_{{\rm out}, J},\varphi_{{\rm down}, J'})=0
\,.
\label{eq:orthogonality}
\end{align}

The ``in" modes undergo scattering by the black hole's potential $V_l$. A portion, $R_J$, is scattered towards $\mathscr{J}^+$, while another portion, $T_J$, penetrates the potential barrier and reaches the horizon $H^+$. The modes going into $\mathscr{J}^+$ and $H^+$ are the ``out" and the ``down" modes, implying that the ``in" mode can be represented by
\begin{equation}
\label{eq:scattering_IN_modes}
    \varphi_{\text{in}, J} = R_J \varphi_{\text{out}, J} + T_J \varphi_{\text{down}, J}
    \,.
\end{equation}
The ``up" modes also scatter off the potential, giving
\begin{equation}
\label{eq:scattering_up_modes}
    \varphi_{\text{up}, J} = t_J \varphi_{\text{out}, J} + r_J \varphi_{\text{down}, J}\,. 
\end{equation}
The orthogonality \eqref{eq:orthogonality} tells us
\begin{equation}
\label{eq:unitary_relation}
|R_J|^2 + |T_J |^2 =1\,, \quad
    |r_J|^2 + |t_J|^2 = 1\,, \quad t_J \bar{R}_J + r_J \bar{T}_J =0\,,
\end{equation}
which yield
\begin{equation}
    |t_J| = |T_J|, \quad |r_J| = |R_J|\,.
    \label{tT_and_rR}
\end{equation}
The complex coefficients $R_J$ and $T_J$ are called reflection and transmission coefficients, characterizing the classical scattering problem in the exterior region (region I) of the black hole, the details of which require solving the radial equation \eqref{eq:radial_equation}.

However, they do not form the complete basis for the treatment of Hawking radiation. Let us introduce the so-called ``dn" mode, which is a mirror-like image of the ``up" mode and ``outgoing'' in region II (see Fig.~\ref{fig:penrose_diagram}). In the Kruskal coordinates $(U,V,\theta,\phi)$ where
\begin{align}
    v=-4GM\ln \frac{V}{4GM},\quad u= - 4GM \ln \frac{-U}{4GM}\,,
\end{align}
the ``dn" mode is defined as 
\begin{align}
    \varphi_{{\rm dn}, J}(U,V,\theta,\phi)=\bar{\varphi}_{{\rm up} , J}(-U,-V,\theta, \phi)
    \,.
\end{align}
The interior of the black hole can now be spanned by the ``dn" and ``down" modes. The necessity of the ``dn" mode can be understood by considering the Vaidya spacetime describing a gravitational collapse, see the right panel of Fig.~\ref{fig:penrose_diagram}. In this case, a wavepacket injected from the past infinity at a certain time can be an outgoing wavepacket inside the black hole (blue line). Note that this picture also explains that the ``up" mode can be identified with the wave injected at the early time. Therefore, if the black hole forms as a result of a gravitational collapse, the ``up" and ``dn" modes are interpreted as the modes injected from the past null infinity, rather than the waves emitted from the horizons. 

Since the gravitational collapse is a dynamical process, a positive frequency state does not remain positive frequency. Hence, we should take certain linear combinations of the modes to correctly identify the modes defined as the positive frequency state at the past null infinity $\mathscr{J}^-$ of the Vaidya spacetime. Such modes are called the ``d" and ``p" modes and given by the following thermal Bogoliubov transformation:\footnote{Note that in the original paper by Hawking \cite{Hawking:1975vcx} the Bogoliubov coefficients are matrix elements between different values of $J$. However, following \cite{Frolov:1998wf} we assume a diagonal form for such interactions. As we will argue later on, this is only related to a normalization in the squeezed $S$-matrix.} 
\begin{equation}
\label{eq:DP_modes}
    \varphi_{{\rm d},J} = c_J \varphi_{{\rm dn},J} + s_J \Bar{\varphi}_{{\rm up},J}, \quad \varphi_{{\rm p},J} = c_J \varphi_{{\rm up},J} + s_J \Bar{\varphi}_{{\rm dn},J}
\end{equation}
where
\begin{equation}
    s_J = \frac{1}{\sqrt{e^{8GM\pi\omega}-1}}, \quad c_J = \frac{e^{4GM\pi\omega}}{\sqrt{e^{8GM\pi\omega}-1}}\,.
    \label{eq:s_c_def}
\end{equation}
Having understood the physical interpretations of the mode functions, we only focus on the Schwarzschild spacetime because we are not interested in the details of the gravitational collapse.

\subsection{$S$-matrix in the Boulware vacuum}\label{subsec:$S$-matrix_in_Boulware_vacuum}

Equipped with the mode functions explained so far, we proceed to define the vacuum state of the Schwarzschild spacetime and the $S$-matrix. Our first choice of the vacuum is the Boulware vacuum, which is the vacuum associated with the static observer in the Schwarzschild spacetime. Denoting the annihilation operators associated with the ``in" and ``out" modes as $\hat{b}_{\text{in}, J}$ and $\hat{b}_{\text{out}, J}$, the Boulware vacuum $\ket{0}$ is defined by
\begin{align}
    \hat{b}_{\text{in},J}\ket{0}=\hat{b}_{\text{out},J}\ket{0}=0
    \,.
\end{align}
Note that the relations \eqref{eq:scattering_IN_modes} and \eqref{eq:scattering_up_modes} do not mix positive and negative frequency modes. Hence, the Boulware vacuum is equally the vacuum of the up and down modes.

Let us focus on the exterior of the black hole. Then, the field operator can be expanded as
\begin{align}
    \hat{\varphi}&=\int_J \hat{\bm{b}}_{\text{in},J} \bm{\Psi}_{\text{in},J} + {\rm h.c.}
    =\int_J \hat{\bm{b}}_{\text{out},J} \bm{\Psi}_{\text{out},J} + {\rm h.c.}
\end{align}
where $\int_J=\sum_{\ell,m}\int\hd \omega /2\omega$ and
\begin{align}
    \bm{\Psi}_{\text{in},J} &= 
    \begin{pmatrix}
        \varphi_{\text{in},J} \\ \varphi_{\text{up},J}
    \end{pmatrix}
    ,\qquad 
    \bm{\Psi}_{\text{out},J} = 
    \begin{pmatrix}
        \varphi_{\text{out},J} \\ \varphi_{\text{down},J}
    \end{pmatrix}
, \\
    \hat{\bm{b}}_{\text{in},J} &=(\hat{b}_{\text{in},J},\hat{b}_{\text{up},J})\,,\qquad \hat{\bm{b}}_{\text{out},J} =(\hat{b}_{\text{out},J},\hat{b}_{\text{down},J})
    \,,
\end{align}
with the commutation relations
\begin{align}
    [b_{\alpha,J},b_{\beta,J'}^{\dagger}]=\delta_{\alpha,\beta}\hdelta_{J,J'}\,, \qquad \alpha,\beta=\text{in, up},
\end{align}
and the similar relations for the out bases. In this matrix notation, the Hermitian conjugation and transposition of matrices are represented by the superscripts of ``$+$'' and ``$~'~$''. One should not confuse ``$\dagger$'' with ``$+$'' which are
\begin{align}
    \hat{\bm{b}}^{\dagger}_{\text{in},J} &=(\hat{b}^{\dagger}_{\text{in},J},\hat{b}^{\dagger}_{\text{up},J})
    \,, \qquad
    \hat{\bm{b}}^{+}_{\text{in},J} =
    \begin{pmatrix} \hat{b}^{\dagger}_{\text{in},J} \\ \hat{b}^{\dagger}_{\text{up},J}
    \end{pmatrix}
    .
    \label{eq:Bogoliubov_b}
\end{align}
The in and out states of the scattering problem in the Boulware vacuum are identified with the states created by $\hat{\bm{b}}_{\text{in},J}^{\dagger}$ and $\hat{\bm{b}}_{\text{out},J}^{\dagger}$ acting on $\ket{0}$, and the generic multi-particle state is denoted by
\begin{align}
    \ket{i_{\text{in},J},j_{\text{up},J}}:=\frac{1}{\sqrt{i!j!}}(\hat{b}^{\dagger}_{\text{in},J})^i (\hat{b}^{\dagger}_{\text{up},J})^j \ket{0}
    \,,
\end{align}
and similar for the out state.

Eqs.~\eqref{eq:scattering_IN_modes} and \eqref{eq:scattering_up_modes} yield
\begin{align}
    \hat{\bm{b}}_{\text{out},J} = \hat{\bm{b}}_{\text{in},J} \bm{T}^+_J
    \,, \qquad
    \bm{T}_J:=\begin{pmatrix}
  \bar{R}_J & \bar{t}_J \\
  \bar{T}_J & \bar{r}_J 
\end{pmatrix}
\,.
\label{bout-bin}
\end{align}
Adopting the matrix notation, the transition probabilities of 1-particle scattering are then computed as
\begin{align}
\braket{\bm{1}_{\text{out},J}}{\bm{1}_{\text{in},J'}}&=\bra{0}\hat{\bm{b}}'_{\text{out},J}\hat{\bm{b}}^{\dagger}_{\text{in},J'}\ket{0}
\nn
&=\bra{0}\hat{\bm{b}}'_{\text{out},J}\hat{\bm{b}}^{\dagger}_{\text{out},J'}{\bm{T}_{J'}^{-1}}'\ket{0}
\nn
&={\bm{T}_{J}^{-1}}'\hdelta_{J,J'}\,,
\end{align}
where ${\bm{T}_{J}^{-1}}'$ is simplified to be
\begin{align}
{\bm{T}_J^{-1}}' =
\begin{pmatrix}
  R_J &  t_J \\
  T_J & r_J 
\end{pmatrix},
\label{eq:smatrix_Boulware}
\end{align}
by using \eqref{eq:unitary_relation}.
We see that the reflection and transmission coefficients, which are scattering amplitudes of classical waves, are identified with the $S$-matrix elements defined in the Boulware vacuum. The generic multi-particle scattering is a tensor product of the 1-particle scattering because the particles are approximated as interacting only with the background.
The $S$-matrix operator, which relate the ``in" and ``out" states via $\hat{\bm{b}}_{\text{in},J}\hat{S}^{\text{B}}=\hat{S}^{\text{B}}\hat{\bm{b}}_{\text{out},J}$ and $\hat{\bm{b}}^{\dagger}_{\text{in},J}\hat{S}^{\text{B}}=\hat{S}^{\text{B}}\hat{\bm{b}}^{\dagger}_{\text{out},J}$, is then found to be
\begin{align}\label{eq:S_matrix_Boulware}
\hat{S}^{\text{B}}&=\prod_J \hat{S}_{J}^{\text{B}}\,, \\
\hat{S}_{J}^{\text{B}}&=N\exp \left[\hat{\bm{b}}_{\text{out},J}^\dagger ({\bm{T}_J^{-1}}'-\bm{I} )\hat{\bm{b}}'_{\text{out},J} \right]\,,
\end{align}
where $N$ represents the normal ordering. Note that the Boulware vacuum is invariant under the $S$-operator
\begin{align}
    \hat{S}^{\text{B}}\ket{0}=\ket{0}
    \,,
\end{align}
because there is no particle production.

\subsection{$S$-matrix in the Unruh vacuum}
\label{subsec:S_matrix_in_Unruh_vacuum}

We next think of a black hole as a result of gravitational collapse. As we examined in Sec.~\ref{sec:mode}, we need the ``dn" mode to form a complete basis in this case. The orthonormal complete bases for the in and out states are known as Wald's bases~\cite{Wald:1975kc}
\begin{equation}
\bm{\Phi}_{\text{in}, J} = \begin{pmatrix}
 \varphi_{\text{in}, J} \\ \varphi_{\text{d}, J} \\ \varphi_{\text{p}, J} 
\end{pmatrix},\  \ \ \ 
\bm{\Phi}_{\text{out}, J} = \begin{pmatrix}
 \varphi_{\text{out}, J} \\ \varphi_{\text{down}, J} \\ \varphi_{\text{dn}, J} 
\end{pmatrix}.
\end{equation}
From these modes, we define the annihilation operators as $\hat{\bm{a}}_{\text{in},J} = (\hat{a}_{\text{in},J}, \hat{a}_{\text{d},J}, \hat{a}_{\text{p},J})$ and $\hat{\bm{a}}_{\text{out},J} = (\hat{a}_{\text{out},J}, \hat{a}_{\text{down},J}, \hat{a}_{\text{dn},J})$, and the in and out vacua by
\begin{align}
    \hat{\bm{a}}_{\text{in},J}\ket{0;{\rm in}}=0\,, \qquad \hat{\bm{a}}_{\text{out},J}\ket{0;{\rm out}}=0
    \,.
\end{align}
The multi-particle states are denoted as
\begin{alignat}{3}
    \ket{i_{\text{in},J},j_{\text{d},J},k_{\text{p},J}}&:=\frac{1}{\sqrt{i!j!k!}}(\hat{a}^{\dagger}_{\text{in},J})^i (\hat{a}^{\dagger}_{\text{d},J})^j (\hat{a}^{\dagger}_{\text{p},J})^k\ket{0;{\rm in}}
    \,,  \\
    \ket{i_{\text{out},J},j_{\text{down},J},k_{\text{dn},J}}&:=\frac{1}{\sqrt{i!j!k!}}(\hat{a}^{\dagger}_{\text{out},J})^i (\hat{a}^{\dagger}_{\text{down},J})^j (\hat{a}^{\dagger}_{\text{dn},J})^k\ket{0;{\rm out}}. 
\end{alignat}
We stress again that the in-vacuum $\ket{0;{\rm in}}$, known as the Unruh vacuum, has been chosen to agree with the in-vacuum of the Vaidya spacetime. The relations~\eqref{eq:scattering_IN_modes}, \eqref{eq:scattering_up_modes}, and \eqref{eq:DP_modes}, are re-organized as
\begin{equation}
\bm{\Phi}_{{\rm in}J}=\bm{A}^+_J\bm{\Phi}_{{\rm out},J} - \bm{B}'_J\Bar{\bm{\Phi}}_{{\rm out},J}
\end{equation}
where
\begin{equation}
\bm{A}_J = 
\begin{pmatrix}
  \Bar{R}_J & 0 & c_J \Bar{t}_J \\
  \Bar{T}_J & 0 & c_J \Bar{r}_J \\
  0 & c_J & 0 
\end{pmatrix}, \ \ \ \ \ 
\bm{B}_J = 
\begin{pmatrix}
  0 & -s_J \Bar{t}_J & 0 \\
  0 & -s_J \Bar{r}_J & 0 \\
  0 & 0 & -s_J
\end{pmatrix}.
\label{eq:Bogolubov_AB}
\end{equation}
The relationship between the annihilation and creation operators is given by
\begin{equation}
\label{eq:relation_anihilation_operator_out_and_in}
\hat{\bm{a}}_{\text{out},J} = \hat{\bm{a}}_{\text{in},J} \bm{A}^+_J - \hat{\bm{a}}^\dagger_{\text{in},J}\bm{B}^+_J.
\end{equation}
In contrast to \eqref{eq:Bogoliubov_b}, the Bogoliubov transformation \eqref{eq:relation_anihilation_operator_out_and_in} mixes the annihilation and creation operators, which results in inequivalence of the in/out vacua and then particle production.

The $S$-matrix operator is now given by\footnote{Notice that the $S$-matrix for the Boulware vacuum (\ref{eq:S_matrix_Boulware}) does not have the $e^{iW^0_J}$ term. Even if introduced, it can easily be seen to be just a phase $e^{iW^0_J}=\theta, |\theta|=1$ since there is no pair production in the Boulware case.} $\hat{S}^\text{U}=\prod_J \hat{S}_{J}^\text{U}$ with~\cite{Frolov:1998wf}
\begin{align}
\label{eq:S_matrix_Unruh}
    \hat{S}_{J}^\text{U} = e^{iW^0_J} N\exp \left[\frac{1}{2} \hat{\bm{a}}_{\text{out},J} \bm{\Lambda}_J \hat{\bm{a}}'_{\text{out},J}
    +\hat{\bm{a}}_{\text{out},J}^\dagger (\bm{M}_J-\bm{I} )\hat{\bm{a}}'_{\text{out},J} 
    + \frac{1}{2} \hat{\bm{a}}_{\text{out},J}^\dagger \bm{V}_J \hat{\bm{a}}'^+_{\text{out},J}\right]
\end{align}
where $N$ denotes the normal ordering operation with respect to the out-state operators, and
\begin{equation}
    \bm{\Lambda}_J = \bm{A}_J^{-1}\bm{B}_J = 
\begin{pmatrix}
  0 & 0 & 0 \\
  0 & 0 & -s_J/c_J \\
  0 & -s_J/c_J & 0 
\end{pmatrix}\,, 
\end{equation}

\begin{equation}
\bm{V}_J = -\Bar{\bm{B}}_J\bm{A}_J^{-1}
=
\begin{pmatrix}
  0 & 0 & s_J t_J/c_J \\
  0 & 0 & s_J r_J/c_J \\
  s_J t_J/c_J & s_J r_J/c_J & 0 
\end{pmatrix}\,,
\label{V_Unruh}
\end{equation}

\begin{equation}
\bm{M}_J = {\bm{A}_J^{-1}}'
=
\begin{pmatrix}
  R_J & 0 & t_J/c_J \\
  T_J & 0 & r_J/c_J \\
  0 & 1/c_J & 0 
\end{pmatrix}\,,  
\end{equation}

\begin{equation}
    e^{iW^0_J} = \theta \left[\det (\bm{A}_J^+ \bm{A}_J)\right]^{-1/4} = \theta c_J^{-1}\,, \quad |\theta| =1\,.
\end{equation}
The derivation of this $S$-matrix is described in Appendix~\ref{sec:Bogoliubov}.
The $S$-operator \eqref{eq:S_matrix_Unruh} involves the squeezing operator due to the non-vanishing $\bm{B}$ matrix. In particular, it implies that the in-vacuum,
\begin{align}
\ket{0;{\rm in}}=\hat{S}^\text{U} \ket{0;{\rm out}}=\prod_J e^{iW^0_J} N\exp \left[ \frac{1}{2} \hat{\bm{a}}_{\text{out},J}^\dagger \bm{V}_J \hat{\bm{a}}'^+_{\text{out},J}\right] \ket{0;{\rm out}}\,,
\label{eq:out_vacuum}
\end{align}
contains the out-state particles always in a pair, as we advertised at the beginning of this section.

\subsection{Hawking radiation as a resummation of emission processes}\label{subsec:Hawking_radiation_as_a_resummation_of_amplitudes}

After having discussed the two choices of vacuum in defining the $S$-matrix on the Schwarzschild black hole, let us derive Hawking's formula for thermal radiation. We consider the Unruh vacuum. Given this, the observable of interest is the number density of particles at $\mathscr{J}_+$:
\begin{align}\label{eq:number}
    n_{\text{out},J}:=\bra{0;{\rm in}}\hat{n}_{\text{out},J}\ket{0;{\rm in}}\,, \qquad \hat{n}_{\text{out},J}:=\hat{a}^{\dagger}_{\text{out},J}\hat{a}_{\text{out},J}
    \,.
\end{align}

The number of particles \eqref{eq:number} can be computed by using the Bogoliubov transformation \eqref{eq:relation_anihilation_operator_out_and_in}. One can rewrite the out-state annihilation and creation operators in terms of the in-state ones. Then, \eqref{eq:number} is given by a component of the Bogoliubov coefficients \eqref{eq:Bogolubov_AB} as
\begin{equation}
\label{eq:Hawking_thermal_distribution_from_in_ev}
     n_{\text{out},J} = s_J^2 |t_J|^2 \hdelta_{J,J}= \frac{|t_J|^2}{e^{8GM \pi \omega}-1} \hdelta_{J,J}\,, 
\end{equation}
which is the well-known Hawking thermal distribution for a black hole. Note that the use of Fourier modes leads to the divergence factor $\hdelta_{J,J}$. In order to obtain a finite quantity, we perform our experiment in a finite time so that the following identity holds
\begin{align}\label{eq:reg-Dirac}
    \hdelta(\omega-\omega)=\int^{1/\omega}_{-1/\omega}\dd t = \frac{1}{2\omega}
    \quad \Rightarrow \quad\hdelta_{J,J}=1\,.
\end{align}
We adopt this prescription in the following calculations.

Having derived (\ref{eq:Hawking_thermal_distribution_from_in_ev}) from the in basis, we revisit the same derivation with the out basis by using the $S$-matrix previously discussed. Using \eqref{eq:out_vacuum}, we find
\begin{align}
    n_{\text{out},J}=\bra{0;{\rm out}}\hat{S}^{\text{U}\dagger}\hat{n}_{\text{out},J}\hat{S}^\text{U} \ket{0;{\rm out}} = \bra{0;{\rm out}}\hat{S}^{\text{U}\dagger}_{J} \hat{n}_{\text{out},J}\hat{S}^\text{U}_{J}\ket{0;{\rm out}}
    \,,
    \label{nout_J}
\end{align}
where we have used unitarity of the $S$-operator to obtain the last expression. Readers familiar with the KMOC formalism~\cite{Kosower:2018adc} will notice that this representation of observables is exactly the same as the KMOC formalism (in the Heisenberg picture). We insert the complete sets of ``out" modes in the full Hilbert spaces $\mathcal{H}_{\text{out}}\otimes \mathcal{H}_{\text{down}}\otimes \mathcal{H}_{\text{dn}}$:\footnote{One can insert the completeness relation of the in state, $n_{\text{out},J}=\sum_{i,j,k}| \bra{i_{\text{in}, J}, j_{\text{d}, J},k_{\text{p}, J}}  \hat{a}_{\text{out}, J} \ket{0;\text{in}}|^2$, which immediately leads to \eqref{eq:Hawking_thermal_distribution_from_in_ev} by using the Bogoliubov transformation, cf.~\cite{Aoki:2024bpj} in the case of cosmological background.}
\begin{align}
\label{eq:Hawking_radiation_KMOC}
n_{\text{out},J} =\sum_{i, j, k}
     | \bra{i_{\text{out}, J}, j_{\text{down}, J},k_{\text{dn}, J}}  \hat{a}_{\text{out}, J} \hat{S}_{J}^\text{U}\ket{0;\text{out}}|^2\,,
\end{align}
where only the fixed $J$ state contributes, so the subscripts $J$ will be omitted in the following.
When the $S$-matrix acts on the out-vacuum $\ket{0;\text{out}}$, the only relevant operator is $\frac{1}{2}\bm{a}^\dagger \bm{V} \bm{a}'^+=\frac{st}{c} \hat{a}^\dagger_{\text{out}} \hat{a}^\dagger_{\text{dn}} + \frac{sr}{c} \hat{a}^\dagger_{\text{down}}\hat{a}^\dagger_{\text{dn}}$. Therefore, after a few algebraic steps, we have 
\begin{align*}
\label{eq:derivation_of_Hawking_radiation1}
&\bra{i_{\text{out}}, j_{\text{down}},k_{\text{dn}}}  \hat{a}_{\text{out}} \hat{S}^\text{U}\ket{0;\text{out}}
\nn
&=\theta c^{-1} \left(\frac{s}{c}\right)^{i+j+1} \frac{t^{i+1}r^j}{(i+1)!j!} \sqrt{(i+1)(i+1)!(i+j+1)!j!}\delta_{k-i-1,j}\,. 
\stepcounter{equation}\tag{\theequation} 
\end{align*}
Then, the expectation value of the number density is calculated as 
\[
\label{eq:derivation_of_Hawking_radiation2}
n_{\rm out}
   &= \sum_{i, j, k} | \bra{i_{\text{out}}, j_{\text{down}},k_{\text{dn}}}  \hat{a}_{\text{out}} \hat{S}^\text{U}\ket{0;\text{out}}|^2
\\&
    =c^{-2} \sum_{i=0}^\infty \sum_{k=i+1}^\infty \left(\frac{s}{c}\right)^{2k} \frac{k!}{i!(k-i-1)!} |t|^{2(i+1)}|r|^{2(k-i-1)}
\\&
    =\frac{s^2|t|^2}{[c^2-s^2(|r|^2+|t|^2)]^2}=\frac{|t|^2}{e^{8GM\pi\omega}-1}\,, 
\]
where we used \eqref{eq:unitary_relation} and \eqref{eq:s_c_def} to obtain the last expression\footnote{Notice that by working in a single-channel basis labeled by 
$J$, we are effectively using diagonal Bogoliubov coefficients. If, instead, we were to follow Hawking's approach, these coefficients would also include off-diagonal terms, leading to apparent divergences in the occupation number. This issue -- well known to Hawking \cite{Hawking:1975vcx} -- has been recently emphasized in \cite{Aoude:2024sve}. However, as discussed in \cite{Frolov:1998wf}, such divergences are absent in that framework.}. Interestingly, we have seen that the Hawking thermal distribution is obtained by resumming all creation processes from the in-vacuum. Although the use of Bogoliubov transformation is more efficient for computing the number density, the advantage of using the $S$-matrix on the background is that we can explicitly see the dynamical process behind the Hawking radiation. In particular, it allows an effective description of the Hawking radiation while treating the black hole as a dynamical object rather than a fixed background, which we discuss in Sec.~\ref{sec:on-shell}.

\subsection{Graviton modes on Schwarzschild spacetime}\label{subsec:Graviton_modes_on_Schwarzschild spacetime}

Having discussed in detail the $S$-matrix on Schwarzschild for scalar modes, we now turn our attention to Hawking radiation of massless fields with spin. We particularly consider gravitational perturbations (gravitons), though the same arguments apply to photons. The new feature of spinning particles is the existence of polarisation degrees of freedom. These are two convenient bases: one is parity basis $P=({\rm RW},{\rm Z})$ 
\begin{align}
    \hat{h}_{\mu\nu}&=\sum_{P={\rm RW,Z}}\int_{J} \hat{\bm{a}}_{\text{in}, \{J,P\}} \left(\bm{\Psi}_{\text{in}, \{J,P\}}\right)_{\mu\nu} + {\rm h.c.}
    =\sum_{P={\rm RW,Z}}\int_{J} \hat{\bm{a}}_{\text{out}, \{J,P\}} \left(\bm{\Psi}_{\text{out}, \{J,P\}}\right)_{\mu\nu} + {\rm h.c.} \,,
\end{align}
and the other is helicity basis $\eta=(+2,-2)$
\begin{align}
    \hat{h}_{\mu\nu}&=\sum_{\eta=\pm 2}\int_{J} \hat{\bm{a}}_{\text{in}, \{J,\eta\}} \left(\bm{\Psi}_{\text{in}, \{J,\eta\}}\right)_{\mu\nu} + {\rm h.c.}
    =\sum_{\eta=\pm 2}\int_{J} \hat{\bm{a}}_{\text{out}, \{J,\eta\}} \left(\bm{\Psi}_{\text{out}, \{J,\eta\}}\right)_{\mu\nu} + {\rm h.c.} \,.
\end{align}
They will approach free waves in the asymptotic region $r\rightarrow +\infty$, where they admit the particle interpretations. The relation between the two bases can thus be fixed as in the Minkowski spacetime~\cite{Thorne:1980ru,Martel:2005ir}
\begin{equation}
\label{eq:creation_annihilation_helicity_parity}
\hat{\bm{a}}_{+2} = \frac{\hat{\bm{a}}_{\rm Z} - i\, \hat{\bm{a}}_{\rm RW}}{\sqrt{2}}, \quad
\hat{\bm{a}}_{-2} = \frac{\hat{\bm{a}}_{\rm Z} + i\, \hat{\bm{a}}_{\rm RW}}{\sqrt{2}}\,.
\end{equation}
where the labels in/out and $J$ are omitted for simplicity of notation.

Thanks to the spherical symmetry of the background, the two distinct types of the parity basis, called axial (odd-parity, or RW for Regge-Wheeler) and polar (even-parity, or Z for Zerilli) modes, are decoupled at the linear order in the BHPT. Focusing on the radial part, the axial perturbation, denoted by $X_{\omega l}$, satisfies the Regge-Wheeler equation~\cite{Regge:1957td}
\begin{equation}
\label{eq:Regge_Wheeler_equation}
        \frac{\dd^2 X_{\omega l}}{\dd r_*^2} + \qty(\omega^2 - V_{l,\text{RW}})X_{\omega l} =0\,, 
\end{equation}
\begin{equation}
\label{eq:Regge_Wheeler_potential}
    V_{l,\text{RW}}(r) =  2\qty(1-\frac{2GM}{r})\frac{nr-3GM}{r^3}\,,
\end{equation}
where $n=l(l+1)/2$. On the other hand, the polar one, denoted by $Z_{\omega l}$, obeys Zerilli's equation~\cite{Zerilli:1970wzz}
\begin{equation}
\label{eq:Zerillis_equation}
    \frac{\dd ^2Z_{\omega l}}{\dd r_*^2} + \qty(\omega^2 -V_{l,\text{Z}})Z_{\omega l} = 0\,,
\end{equation}
\begin{equation}
\label{eq:Zerillis_potential}
    V_{l,\text{Z}} (r) = 2 \qty(1-\frac{2GM}{r}) \frac{n^2 (n+1) r^3 + 3n^2 G M r^2 + 9nG^2M^2 r + 9G^3 M^3}{r^3 (nr + 3GM)^2}\,. 
\end{equation}

Despite the different forms of these differential equations, they are related through specific differential operations~\cite{Anderson:1991kx}. Concretely, the two master variables are related by
\begin{equation}
    \qty[\frac{2}{3}n(n+1)-2GMi\omega]X_{\omega l} = \qty[\frac{2}{3}n(n+1) + \frac{6G^2 M^2(r-2GM)}{r^2(nr+3GM)}]Z_{\omega l} -2GM \frac{\dd Z_{\omega l}}{\dd r_*}\,. 
\end{equation}
Therefore, a solution of Zerilli's equation~\eqref{eq:Zerillis_equation} under the following boundary condition
\begin{equation}
    Z_{\omega l}\to
    \begin{cases}
        e^{-i\omega r_*} + R_{\omega l,\text{Z}} e^{+i\omega r_*} & (r_*\to +\infty)\\
        T_{\omega l,\text{Z}} e^{-i\omega r_*} & (r_*\to -\infty)\,,
    \end{cases}
\end{equation}
is equivalent to the solution of the Regge-Wheeler equation~\eqref{eq:Regge_Wheeler_equation} with the following boundary condition~\cite{Chandrasekhar:1975zza}

\begin{equation}
\label{eq:boudary_condition_Regge_Wheeler_Chandrasekhar}
    X_{\omega l} \to
    \begin{cases}
        e^{-i\omega r_*} + \underset{=:R_{\omega l,\text{RW}}}{\underbrace{\frac{\frac{2}{3}n(n+1) + 2GMi\omega}{\frac{2}{3}n(n+1) - 2GMi\omega} R_{\omega l, \text{Z}}}} e^{+i\omega r_*} & (r_*\to +\infty)\\
        \underset{=:T_{\omega l,\text{RW}}}{\underbrace{T_{\omega l, \text{Z}}}} e^{-i\omega r_*} &(r_*\to -\infty)\,.
    \end{cases}
\end{equation}
Consequently, the transmission coefficients for axial and polar perturbations are identical, while the reflection coefficients agree up to a phase shift. In particular, the magnitudes of the reflection and transmission amplitudes are equal for both types of perturbations: 
\begin{equation}
\label{eq:relection_transmission_coefficients_btw_RW_Z}
|R_{\omega l,\text{RW}}|=|R_{\omega l, \text{Z}}|\,, \quad
T_{\omega l,\text{RW}} = T_{\omega l,\text{Z}}~. 
\end{equation}

Building upon the preceding discussion, we construct the $S$-matrix for graviton perturbations. With the vacuum state defined via the annihilation operators, the Hilbert space at each $J$ can be thought of as the tensor product of the parity sectors $\mathcal{H}_{\text{RW}} \otimes\mathcal{H}_{\text{Z}}$. These two sectors are decoupled within the linear BHPT. Therefore, the $S$-matrix of each $J$ is factorized into the distinct parity sectors
\begin{align}
    \hat{S}_{J} =\hat{S}_{J,\text{RW}}\hat{S}_{J,\text{Z}}  \,.
\end{align}
The same factorization holds regardless of the choice of the Boulware or the Unruh vacuum, so we do not write the labels B or U. The full $S$-matrix is thus
\begin{align}
    \hat{S}=\prod_{J,P}\hat{S}_{J,P}
    \,.
\end{align}
The $S$-matrix of each $J$ and $P$ is constructed from the transmission and reflection coefficients in the same way as the scalar case. Having obtained the $S$-matrix on the parity basis, the $S$-matrix on the helicity basis can be obtained by applying the transformation \eqref{eq:creation_annihilation_helicity_parity}.

\section{On-shell approach to black holes}
\label{sec:on-shell}

In the previous section, we studied semi-classical scattering problems where the black hole is treated as a classical fixed background and only the perturbations are quantized. In this section, we look at the same problems from a different perspective by treating a black hole as a ``particle''. More specifically, we treat a black hole as an on-shell one-particle state $\ket{p,s,A}$ labelled by the momentum $p$ with $p^2=\mu^2$, the spin $s$, and possible other quantum numbers $A$ representing microscopic degrees of freedom of a black hole, with the completeness relation
\begin{align}\label{eq:spec-dens}
    \hat{1}_{\rm BH}=\sum_{s,A}\int \dd \mu^2\, \int \dd\Phi(p)\rho_{s,A}(\mu^2)\ket{p,s,A}\bra{p,s,A}
    \,,
\end{align}
where $\rho_{s,A}(\mu^2)$ is the spectral density.\footnote{A single particle state corresponds to $\rho(\mu^2)= \:\delta(\mu^2-M^2)$, which corresponds to an object with a definite mass and spin, and commonly used in the Post-Minkowskian approximation. In the present paper, $\rho_{s,A}(\mu^2)$ is a continuum spectral allowing the change, so the state is different from the standard notion of the single particle state.} We then describe processes of black holes by using the on-shell $S$-matrix. This approach is inspired by the recent programme of applying scattering amplitudes to classical gravitational physics. Indeed, this approach has been successfully employed to describe classical horizon absorptions~\cite{Aoude:2023fdm, Jones:2023ugm, Chen:2023qzo, Bautista:2024emt} and a merger of classical black holes \cite{Aoki:2024boe}. We would like to extend this programme to accommodate the quantum effects of black holes, such as the Hawking radiation. For simplicity of presentation, we only study the absorption and emission of gravitons as the scalar case is simply obtained by neglecting the polarisation degrees of freedom.

\subsection{Amplitudes for black hole absorption and emission}

We start by elaborating on the basic idea of the on-shell approach to black holes. The key ingredient is a mass-changing amplitude \cite{Aoude:2023fdm, Jones:2023ugm}: whether or not the process is classical or quantum, the mass of a black hole has to change when it absorbs or emits a particle according to the conservation laws. In general, the spin and other quantum numbers can also change. The simplest process is the one-particle absorption/emission, which is described by the mass-changing three-point amplitude~\cite{Arkani-Hamed:2017jhn}
\begin{align}
\label{eq:3-point-EFT}
\begin{tikzpicture}[baseline=-2]
\begin{feynhand}
\propag (1.4, -1) node[right] {$p$} -- (0,0) ;
\draw[very thick] (0,0) -- (-1.4, 0) node [left] {$p_X, s$};
\propag [boson]  (0,0) -- (1.4, 1) node [right] {$k,\eta$};
\vertex[grayblob] at (0,0){};
\end{feynhand}
\end{tikzpicture}
:~
\mathcal{A}_3=g M^{1-2s}\langle{\bm p}_X k\rangle^{s-\eta}[{\bm p}_X k]^{s+\eta} 
\,,
\end{align}
where the solid line is a Schwarzschild black hole of the mass $M$ and the wavy line is the absorbed $(k^0>0)$ or emitted $(k^0<0)$ graviton with the helicity $\eta$.\footnote{Precisely speaking, we should use a wavepacket state to describe a semi-classical object~\cite{Kosower:2018adc}. We will discuss this in Sec.~\ref{sec:observables}.}  The bold line is the mass-changed black hole state of spin $s$, which we call the $X$ state. According to the conservation, the mass of the $X$ state is $\mu^2=M^2+2M\omega$ where $\omega:=u\cdot k$ with $u^{\mu}:=p^{\mu}/M$ being the four-velocity. Here, we have adopted the spinor-helicity formalism, especially the boldface notation for the little group indices of massive spinning particles~\cite{Arkani-Hamed:2017jhn}. The kinematic dependence of this amplitude is uniquely fixed, and the only free parameter is the overall coupling $g$. Note that this coupling need not be elementary: it can depend on the external states, i.e.,~the helicity of the graviton,  masses and spins of the black holes, and possibly their microscopic quantum numbers.

The on-shell approach here is an effective theory in the sense that it requires a number of coupling constants and the spectral density as inputs. These parameters are fixed to recover the predictions of a ``fundamental'' theory. The advantage of the effective theory is to provide a universal description of phenomena. For instance, while we take the fundamental theory as the linear BHPT, one can apply similar arguments in quantum gravity, in principle. The difference in the underlying theories is encoded in the difference in the parameters of the effective theory. As a concrete example, we will see that the difference in the vacuum choice is captured by the different choice of a certain combination of the spectral density and the couplings. Furthermore, one can always recycle computations into different setups if need be. In our context, we determine the couplings from the data of a single black hole. We can then use them to compute observables in binary black holes. In this way, one can efficiently compute the observables in the binary system from BHPT~\cite{Jones:2023ugm, Bautista:2024emt}.

Therefore, our task is to provide a concrete dictionary between the fundamental theory, which is the linear BHPT in our context, and the on-shell description of the black holes, especially after integrating out unnecessary degrees of freedom for computing observables. For this purpose, we should first give a brief sketch of how observables are computed in our approach. We employ the KMOC formalism~\cite{Kosower:2018adc} developed in the studies of gravitational waves. While the main point of~\cite{Kosower:2018adc} is a computation of (semi-)classical dynamics, the foundation of the KMOC formalism is quite generic, and it can be applied to our context. Indeed, a related semi-classical application of KMOC  was recently provided in  \cite{Aoude:2024sve}, where the amplitudes eikonal  was employed to derive the Hawking spectrum. 

Let us now work in the interaction picture. We prepare an initial state $\ket{\Psi}$ with $\braket{\Psi}=1$ and make an experiment after time evolution. Since observables are expectation values and the time evolution operator is given by the $S$-matrix, the observable associated with the operator $\hat{O}$ is given by
\begin{align}
    O=\bra{\Psi}\hat{S}^{\dagger}\hat{O}\hat{S}\ket{\Psi}\,.
    \label{Observable}
\end{align}
By representing $\hat{O}$ by using the annihilation and creation operators, we can insert the completeness relation into \eqref{Observable} to express it as a square of the $S$-matrix elements. For example, the number operator $\hat{n}_{J}=\hat{a}_{J}^{\dagger}\hat{a}_{J}$ gives
\begin{align}\label{eq:density-operator}
n_{J}=\sum_n\bra{\Psi}\hat{S}^{\dagger}\hat{a}_{J}^{\dagger}\ketbra{n}\hat{a}_{J}\hat{S}\ket{\Psi}\,,
\end{align}
which we computed in Sec.~\ref{subsec:Hawking_radiation_as_a_resummation_of_amplitudes} within the BHPT (in the Heisenberg picture).
We stress that this type of observable is inclusive because we should sum over all possible final states. This connects the idea of effective theories. When one is interested in observables of a certain sector (visible sector), we can first perform the summation over its complement (invisible sector) to formulate an effective description. Said another way, the matching between the fundamental and effective theories is conducted through inclusive observables summing over the invisible sector.

While one such observable is the absorption cross-section~\cite{Aoude:2023fdm, Jones:2023ugm, Chen:2023qzo, Aoki:2024boe}, in the present paper, we use the discontinuity of the absorptive part of the graviton scattering ($\text{BH} + \text{graviton} \to \text{BH} + \text{graviton}$). The advantage is that it can be applied to helicity-violating processes~\cite{Bautista:2024emt}. First, the four-point scattering amplitude of a Schwarzschild black hole and a graviton is denoted by
\begin{equation}
     \bra{p'; k',\eta'} \hat{T} \ket{p; k,\eta}=\hat{\delta}^4( p+k-p'-k')\mathcal{A}^{-\eta', \eta}_{4}\,.
\end{equation}
Throughout the present paper, we use the conventional out-in bra-ket notation. In the spinor-helicity formalism, on the other hand, it is customary to use the all-ingoing (or all-outgoing) notation; indeed, the helicity sign in \eqref{eq:3-point-EFT} is defined for the ingoing momentum. Hence, we adopt the all-ingoing helicity notation in the amplitudes, and the helicity label of the out state is opposite to that in the bra. 
The amplitudes $\Amp_4^{+,-}, \Amp_4^{-,+}$ are helicity-preserving whereas $\Amp_4^{+,+}, \Amp_4^{-,-}$ are helicity-violating. According to unitarity, the discontinuity of the amplitude can be expressed by a sum of the on-shell cuts of internal states. We specifically consider the cut of the internal $X$ state:\footnote{Given a function of complex variables $f(z)$, we define its discontinuity as $
    {\rm Disc}f(z) \coloneqq f(z +i\epsilon)- f(z -i\epsilon)$. The subscript/superscript records that we compute the discontinuity across cuts associated with particular states.}
\begin{equation}
\label{eq:discontinuity_4point_amplitude}
\begin{tikzpicture}[baseline=-2]
\begin{feynhand}
\propag (1.6, -0.6) node[right] {$p$} -- (0.7,0) ;
\propag [boson]  (0.7,0) -- (1.6, 0.6) node [right] {$k,\eta$};
\draw[very thick] (0.7,0) -- (-0.7, 0);
\propag (-1.6, -0.6) node[left] {$p'$} -- (-0.7,0) ;
\propag [boson]  (-0.7,0) -- (-1.6, 0.6) node [left] {$k',\eta'$};
\draw[red, dashed] (0, 0.4) node[above] {$p_X,s,A$} -- (0, -0.4) ;
\vertex[grayblob] at (0.7,0){};
\vertex[grayblob] at (-0.7,0){};
\end{feynhand}
\end{tikzpicture}
:~{\rm Disc}_X \mathcal{A}_{4}^{\eta',\eta}={ 2\pi i} \sum_{s,A} (\mathcal{A}_{3,A}^{-\eta', s})^* \mathcal{A}_{3,A}^{\eta,s} \, \rho_{s,A}(\mu^2)
\,,
\end{equation}
where the contraction of the little group indices is understood. The absorption/emission of the $l$-mode of a graviton correspnds to the internal spin-$s$ state,\footnote{It follows the fact that the kinematic dependence of the three-point amplitudes is equal to spin-weighted spherical harmonics, see e.g.,~\cite{Aoude:2023fdm}.}
\begin{align}
    {\rm Disc}_X^{s} \mathcal{A}_{4}^{\eta',\eta}&={16\pi i (2s+1)} \varrho^{s}_{\eta',\eta} \,\mathcal{P}^{s}_{\eta',\eta}
    \,,
    \label{eq:disc_eq_varrho_P}
\end{align}
where $\varrho^s_{\eta',\eta}$ and $\mathcal{P}^s_{\eta',\eta}$ are the discontinuity of the partial-wave amplitudes and the partial-wave basis.\footnote{We will immediately find $\varrho^s_{+,-}=\varrho^s_{-,+}$ and $\varrho^s_{+,+}=\varrho^s_{-,-}$, but we keep them independent for a while to see how they arise in the language of black holes.} While the generic $l$ extension is straightforward, we focus on the spin-2 mode for which they are given by 
\begin{alignat}{3}
    \varrho^{s=2}_{+,-}&:=\frac{2\omega^4}{5M^2}\sum_A g^*_{-,A}g_{-,A}\rho_A\,, \qquad& 
    \varrho^{s=2}_{-,+}&:=\frac{2\omega^4}{5M^2}\sum_A g^*_{+,A}g_{+,A}\rho_A
    \,, \label{rho_preserving} \\
    \varrho^{s=2}_{+,+}&:=\frac{2\omega^4 \mu^4}{5M^6}\sum_A g^*_{-,A}g_{+,A}\rho_A\,, \qquad &
    \varrho^{s=2}_{-,-}&:=\frac{2\omega^4\mu^4 }{5M^6}\sum_A g^*_{+,A}g_{-,A}\rho_A
    \,, \label{rho_violating}
\end{alignat}
(above we have suppressed $s=2$ labels in the r.h.s.) and
\begin{align}\label{eq:disc-struc_preserve}
    \mathcal{P}^{s=2}_{+,-}&:= \frac{[k'|u|k\rangle^4}{16\omega^4} 
    \,, \quad \mathcal{P}^{s=2}_{-,+}:=\frac{\langle k'|u|k]^4 }{16\omega^4}\,,
    \\
\label{eq:disc-struc_violate}   \mathcal{P}^{s=2}_{+,+}&:=\frac{[k'k]^4 }{16\omega^4}\,, \qquad \mathcal{P}^{s=2}_{-,-}:=\frac{\langle k'k \rangle^4}{16\omega^4}
    \,.
\end{align}
Recall that $u^\mu:=p^\mu/M$ is the four-velocity and $\omega:=u\cdot k$ is the frequency of the graviton in the rest frame of $p$. The functions $\mathcal{P}^{s=2}_{\eta'\eta}$ are chosen to agree with the Wigner-d function $d^j_{h',h}$ with $j=2$ and $|h'|=|h|=2$: for example, taking the rest frame $u=(1,\bm{0})$, we find $\mathcal{P}^{s=2}_{+,-}=d^{j=2}_{-2,-2}=\cos^4(\theta/2)$ where $\theta$ is the scattering angle between $k$ and $k'$. The partial-wave discontinuities $\varrho^{s=2}_{\eta',\eta}$ are defined after summing over the invisible degrees of freedom $A$. The discussion can be easily extended into the generic $s$. The discontinuities $\varrho^{s}_{\eta',\eta}$ are what we wish to determine through the matching to a fundamental theory.

In the forward limit, the discontinuity is represented by the cross-section
\begin{align}
\frac{1}{4M\omega i}{\rm Disc}_X \Amp^{-\eta  ,\eta}_4|_{k'\to k}= {}_{\rm abs}\sigma_{\eta} = \sum_{l}{}_{\rm abs}\sigma^{l}_{\eta} 
\,,
\end{align}
where $_{\rm abs}\sigma^{l}_{\eta}$ is the absorption cross-section of the $l$-mode of a graviton with the helicity $\eta$. 
The cut of each spin-$s$ state is the  absorption cross-section for the $l=s$ mode

\begin{align}
\label{mass-changing crosssection}
\frac{20\pi }{ M \omega }\varrho^{s=2}_{-,+}=   {}_{\rm abs}\sigma^{l=2}_+\,, \qquad  
   \frac{20\pi}{M \omega }\varrho^{s=2}_{+,-}={} _{\rm abs}\sigma^{l=2}_{-}\, .
\end{align}
In this way, we can determine the helicity-preserving sectors $\varrho^{s}_{-,+}$ and $\varrho^s_{+,-}$ from the absorption cross-section~\cite{Aoude:2023fdm, Jones:2023ugm, Chen:2023qzo}. However, this cannot be applied to the helicity-violating ones $\varrho^s_{+,+}$ and $\varrho^s_{-,-}$. As we have seen and as studied in \cite{Bautista:2024emt}, they can be fixed by the cut of the four-point amplitude. Nonetheless, it is not so straightforward to isolate the absorptive part from the four-point amplitude beyond the leading order in the small $\omega$ expansion~\cite{Bautista:2024emt}. Our approach, on the other hand, is simple: we compute the summation over $A$ in \eqref{rho_preserving} and \eqref{rho_violating}. We can do this because we know the complete $S$-matrix in the linear BHPT from Sec.~\ref{sec:Quantum_fields_in_Schwarzschild}. This also clarifies what the $X$ state is, and provides a more concrete relation between the BHPT and the on-shell approach.

\subsection{Matching to the Boulware vacuum}

Let us start with our matching procedure for the Boulware vacuum. For simplicity, we focus exclusively on the $l=2$ modes of perturbations, corresponding to the spin-2 $X$-state. To simplify the notation, we do not write the label $J$ and only keep the parity or helicity label.

The $S$-matrix in Sec.~\ref{subsec:$S$-matrix_in_Boulware_vacuum} describes transition probabilities of quanta on the fixed background. On the other hand, we also want to treat a black hole as a quantum state in the on-shell approach. Let us first resolve this discrepancy by following~\cite{Aoki:2024boe}. The ``in/out" modes of the BHPT admit the particle interpretations at $\mathscr{I}^{\mp}$. Hence, combining with the background black hole, we can interpret $\ket{1_{\text{in},P},0_{\text{up},P}}$ and $\ket{1_{\text{out},P},0_{\text{down},P}}$ as the two-particle in and out states in the on-shell approach. Recall that $P=(\text{RW},\text{Z})$ stands for parity-odd or even as discussed in \S~\ref{subsec:Graviton_modes_on_Schwarzschild spacetime}. In contrast, the ``up/down" modes are defined on the horizon. While their particle interpretations would be obscure on their own, we can interpret them as composite particles composed of the background and emitted/injected quanta. The $X$ states in the on-shell approach should thus be matched to these composite states (see Fig.~\ref{fig:composite_particle}). They carry quantum numbers $A$ arising from the degrees of freedom of injected quanta, e.g., the parity degrees of freedom. Note that we are interested in black holes, rather than white holes; thus, we consider only the ``down" modes hereafter.

\begin{figure}[]
 \centering
 \includegraphics[width=0.5\textwidth]{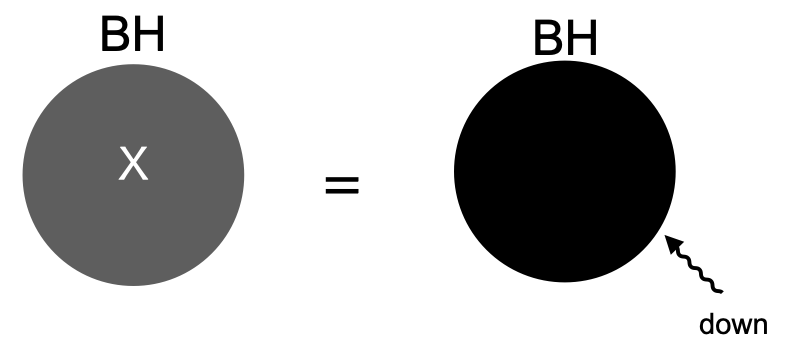}
 \caption{The black hole $X$-state as a composite particle made of the background black hole and the injected quanta in the Boulware vacuum.}
 \label{fig:composite_particle}
\end{figure}

In the case of the Boulware vacuum, we use (the graviton extension of) the $S$-matrix~\eqref{eq:S_matrix_Boulware} as the predictions of our fundamental theory. The state $\ket{1_{\text{in},P},0_{\text{up},P}}$ evolves either the scattered state $\ket{1_{\text{out},P},0_{\text{down},P}}$ or the composite state $\ket{0_{\text{out},P},1_{\text{down},P}}$. We are interested in the latter. The partial-wave discontinuities in the parity basis can thus be computed by
\begin{align}
    \sum_{P''}\bra{0_{{\rm out},P''},1_{{\rm down},P''}}\ket{1_{{\rm in},P'},0_{{\rm up},P'}}^*\bra{0_{{\rm out},P''},1_{{\rm down},P''}}\ket{1_{{\rm in},P},0_{{\rm up},P}}
    \,.
\end{align}
Since the even and odd modes are decoupled, the non-diagonal components vanish. The diagonal components are nothing but the absorption cross-section for the $l=2$ modes
\begin{align}
\label{eq:Boulware_cross_section}
_{\rm abs}\sigma^{\rm B}_{P}(\omega):=\frac{5\pi}{\omega^2} \left|\bra{0_{{\rm out},P},1_{{\rm down},P}}\ket{1_{{\rm in},P},0_{{\rm up},P}}\right|^2 =\frac{5\pi}{\omega^2}|T_{P}|^2\,.
\end{align}
On the other hand, we would like to determine the partial-wave discontinuities in the helicity basis. The relation between the two bases follows from \eqref{eq:creation_annihilation_helicity_parity}. As a result, we find
\begin{align}\label{eq:Boulware-rho}
    \varrho^{\rm B}_{+,-}(\mu^2)
    &=\varrho^{\rm B}_{-,+}(\mu^2)
    = 
    \frac{M\omega}{40\pi} ( {}_{\rm abs}\sigma^{\rm B}_{\text{Z}}(\omega)+ {}_{\rm abs}\sigma^{\rm B}_{\text{RW}}(\omega)) =\frac{M}{4\omega}|T(\omega)|^2
    \,, \\
    \varrho^{\rm B}_{+,+}(\mu^2)
    &=\varrho^{\rm B}_{-,-}(\mu^2)
    =
     \frac{M\omega}{40\pi} ( {}_{\rm abs}\sigma^{\rm B}_{\text{Z}}(\omega) - {}_{\rm abs}\sigma^{\rm B}_{\text{RW}}(\omega))=0
    \,,
\end{align}
for $\omega >0$ with $\mu^2=M^2+2M\omega$. Here, the spherical symmetry (the decoupling between $P={\rm RW}, {\rm Z}$ modes) leads to $\varrho^{\rm B}_{+,-}=\varrho^{\rm B}_{-,+}$ and $\varrho^{\rm B}_{+,+}=\varrho^{\rm B}_{-,-}$. Then, the duality relation \eqref{eq:relection_transmission_coefficients_btw_RW_Z}, especially $|T_{\text{Z}}|=|T_{\text{RW}}|=:|T|$, gives the last equality of each equation. The helicity-preserving sectors are fixed by the greybody factor for graviton scattering $|T|^2$, and the helicity-violating sectors vanish due to the duality of two parity sectors~\cite{Chandrasekhar:1975zza,Teukolsky:1973ha,Porto:2007qi}.

Black holes in the Boulware vacuum do not evaporate. Therefore, the discontinuities are non-vanishing only for $\omega >0$, resulting in absorptive effects only.\footnote{One should not be confused with the process $\bra{1_{{\rm out},P},0_{{\rm down},P}}\ket{0_{{\rm in},P},1_{{\rm up},P}}$, which should be understood as an emission from a white hole.} Thus, despite being matched to a quantum theory, in this sense, the black holes in the Boulware vacuum are ``classical''. They would be enough to describe classical absorption effects of black holes as already used in the literature~\cite{Aoude:2023fdm, Jones:2023ugm, Chen:2023qzo, Bautista:2024emt, Aoki:2024boe}. However, things will change when we consider the Unruh vacuum, where the Hawking radiation comes in.

Before moving to the Unruh vacuum, we provide more concrete expressions of the partial-wave discontinuities. The low-frequency expansion of the greybody factor for the $l=2$ modes of gravitons is~\cite{Starobinsky:1973aij,Starobinskil:1974nkd,Page:1976df,Poisson:1994yf}
\begin{align}
    |T|^2= \frac{4}{225}(2GM \omega)^6\left(1+2\pi GM\omega + \mathcal{O} (\omega^2)\right)
    \,,
\end{align}
which corresponds to
\begin{align}
\label{eq:Boulware_cross_section_aftermatching}
_{\rm abs}\sigma^{\rm B}_{\rm RW}(\omega)={}_{\rm abs}\sigma^{\rm B}_{\rm Z}(\omega)=\frac{4\pi}{45 \omega^2} (2GM\omega)^{6} \left(1+2\pi GM\omega + \mathcal{O} (\omega^2)\right).
\end{align}
For later convenience, we keep the expression up to the sub-leading order. As a result, the low-frequency expansion of $\varrho^{\rm B}_{+,-}$ is given  by
\begin{align}
    \varrho^{\rm B}_{+,-}=\frac{M}{225 \omega} (2GM\omega)^{6} \left(1+2\pi GM\omega + \mathcal{O} (\omega^2)\right)
    \,.
\end{align}

\subsection{Matching to the Unruh vacuum}
\label{sec:matching_Unruh}
Choosing the Unruh vacuum, the $S$-matrix \eqref{eq:S_matrix_Unruh} has a different structure compared to the Boulware case due to the existence of pair productions. The matrix \eqref{V_Unruh} possesses two distinct processes, the ``out-dn" pair production and the ``down-dn" pair production. The ``out" mode is the graviton mode at $\mathscr{I}^+$. Together with the background black hole, the ``down" and ``dn" modes are respectively regarded as the composite $X$ states with increased and decreased masses. Hence, the ``out-dn" pair production is regarded as a decay process of the black hole. On the other hand, the ``down-dn" pair production can be interpreted as a rearrangement of the internal structure of the black hole since the net mass of the black hole does not change (See Fig.~\ref{fig:xstate_Unruh}).
\begin{figure}[t]
 \centering
 \includegraphics[width=\textwidth]{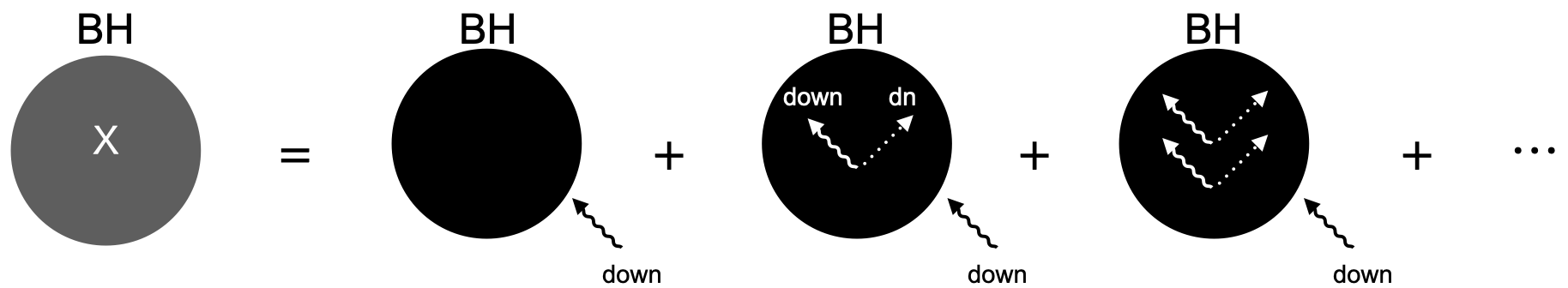}
 \caption{The black hole $X$-state as a composite particle formed by the background black hole together with the injected and pair-created (down-dn) quanta in the Unruh vacuum. Since down–dn pair production preserves the mass while modifying the internal structure, multiple such pair creations are included in the $X$-state.}
 \label{fig:xstate_Unruh}
\end{figure}
The latter processes have to be taken into account in the matching because $\varrho_{\eta'\eta}$ is inclusive, requiring a summation over all invisible processes.

Let us elaborate on the matching conditions in the Unruh vacuum. For $\omega >0$, the discontinuities are matched to absorption processes. As in the Boulware vacuum, it is sufficient to compute the $l=2$ partial-wave absorption cross-sections
\begin{align}
    _{\rm abs}\sigma^{\rm U}_{P}(\omega)&:=\frac{5\pi}{\omega^2}\sum_{m,n}\left|\bra{0_{{\rm out},P},m_{{\rm down},P},n_{{\rm dn},P}}\ket{1_{{\rm in},P},0_{{\rm d},P},0_{{\rm p},P}}\right|^2 \nonumber\\
    &\,=\frac{5\pi}{\omega^2}\frac{|T_{P}|^2{ (1-e^{-8\pi GM\omega})}}{\left(1-|r_{P}|^2e^{-8\pi GM\omega}\right)^2}\,.
\end{align}
Black holes in the Unruh vacuum also exhibit emission processes, which yield non-zero values of discontinuities even for $\omega <0$.
We define
\begin{align}
\label{eq:emission-cross-section}
    _{\rm em}\sigma^{\rm U}_{P}(\omega)
    &:=\frac{5\pi}{\omega^2}\sum_{m,n}\left|\bra{1_{{\rm out},P},m_{{\rm down},P},n_{{\rm dn},P}}\ket{0_{{\rm in},P},0_{{\rm d},P},0_{{\rm p},P}}\right|^2 \nonumber \\
    &\,=\frac{5\pi}{\omega^2}\frac{|t_{P}|^2e^{-8\pi GM\omega}(1-e^{-8\pi GM\omega})}{\left(1-|r_{P}|^2e^{-8\pi GM\omega}\right)^2}\,,
\end{align}
which denotes the decay probabilities normalised to be the same dimension as the cross-section. These results are in agreement with the calculations of Refs.~\cite{Bekenstein:1975tw,Bekenstein:1977mv,Panangaden:1977pc}, which were derived from entropic arguments for black holes. The frequency $\omega $ in \eqref{eq:emission-cross-section} is $\omega >0$ because the graviton in the bra is outgoing. The sign of $\omega$ and helicity must be flipped when it is matched to the amplitudes with the ingoing notation. Note that the ``d" and ``p" modes are regarded as the injected quanta before the formation of the black hole. Since we are interested in processes after the formation, we only consider the ``in" mode as the initial state. We also note that, while pair productions of particles with different quantum numbers $J'\neq J, P'\neq P$ occur, these processes do not contribute to the partial-wave discontinuities of $J$ and $P$ thanks to the unitarity of the $S$-matrix (cf.~\eqref{nout_J}). All in all, the matching conditions in the Unruh vacuum are
\begin{align}\label{eq:Unruh-rho_helicity_perserving}
    \varrho^{\rm U}_{+,-}=\varrho^{\rm U}_{-,+}=
    \begin{dcases}
        \begin{aligned}
        &\frac{M\omega}{40\pi} ( {}_{\rm abs}\sigma^{\rm U}_{\text{Z}}(\omega)+ {}_{\rm abs}\sigma^{\rm U}_{\text{RW}}(\omega)) \\
        &\qquad=\frac{M}{4\omega}\frac{|T(\omega)|^2(1-e^{-8\pi GM\omega})}{\left(1-|R(\omega)|^2e^{-8\pi GM\omega}\right)^2}\,, 
        \end{aligned} 
        &(\omega>0),
        \\
        \begin{aligned}
        &\frac{M|\omega|}{40\pi} ( {}_{\rm em}\sigma^{\rm U}_{\text{Z}}(-\omega)+ {}_{\rm em}\sigma^{\rm U}_{\text{RW}}(-\omega))
        \\
        &\qquad =
        \frac{M}{4|\omega|}\frac{|T(-\omega)|^2e^{-8\pi GM|\omega|}(1-e^{-8\pi GM|\omega|})}{\left(1-|R(-\omega)|^2e^{-8\pi GM|\omega|}\right)^2}\,,
                \end{aligned}
        &(\omega<0),
    \end{dcases}
\end{align}
and
\begin{align}\label{eq:Unruh-rho_helicity_violating}
    \varrho^{\rm U}_{+,+}=\varrho^{\rm U}_{-,-}=
    \begin{dcases}
        \frac{M\omega}{40\pi} ( {}_{\rm abs}\sigma^{\rm U}_{\text{Z}}(\omega)- {}_{\rm abs}\sigma^{\rm U}_{\text{RW}}(\omega)) =0\,, &(\omega>0),
        \\
        \frac{M|\omega|}{40\pi} ( {}_{\rm em}\sigma^{\rm U}_{\text{Z}}(-\omega)- {}_{\rm em}\sigma^{\rm U}_{\text{RW}}(-\omega))=
        0\,,
        &(\omega<0),
    \end{dcases}
\end{align}
where we have used \eqref{tT_and_rR} and \eqref{eq:relection_transmission_coefficients_btw_RW_Z}. 

In contrast to the Boulware vacuum, black holes in the Unruh vacuum have multi-particle processes, e.g.,~$\bra{2_{{\rm out},P},m_{{\rm down},P},n_{{\rm dn},P}}\ket{0_{{\rm in},P},0_{{\rm d},P},0_{{\rm p},P}}$ which should be matched to the four-point amplitude (BH $\to X$ + graviton + graviton). The three-point amplitudes \eqref{eq:3-point-EFT} are not enough to capture the complete dynamics of the linear BHPT in the Unruh vacuum. However, the contributions from the multi-particle processes to observables are well suppressed as we will show in Sec.~\ref{sec:on-shell_Hawking}. Therefore, we only consider the three-point amplitudes in the following.

It is interesting to look at the low-frequency region. The low-frequency expansion of the absorption cross-section at $l=2$ is expressed as
\begin{align}
    _{\rm abs}\sigma^{\rm U}_{\rm RW}(\omega)={}_{\rm abs}\sigma^{\rm U}_{\rm Z}(\omega)=\frac{1}{45\omega^2} (2GM\omega)^5 \left(1+ 6\pi GM\omega +\mathcal{O}(\omega ^2)\right)~,
    \label{low_sigma_U}
\end{align}
while the emission is
\begin{align}
    _{\rm em}\sigma^{\rm U}_{\rm RW}(\omega)={}_{\rm em}\sigma^{\rm U}_{\rm Z}(\omega)=\frac{1}{45\omega^2} (2GM\omega)^5 \left(1- 2\pi GM\omega +\mathcal{O}(\omega ^2)\right) \,.
\end{align}
Comparing \eqref{low_sigma_U} with the absorption cross-section of the Boulware vacuum \eqref{eq:Boulware_cross_section_aftermatching}, the Unruh absorption cross-section is enhanced by $(GM\omega)^{-1}$, which is independent of $\hbar$ when reintroduced. It appears to contradict the intuition that the Hawking effect is inherently quantum. The difference in the choice of vacua would be irrelevant when we take the classical limit $\hbar \to 0$. As pointed out in~\cite{Goldberger:2019sya,Goldberger:2020geb} in the context of worldline effective theory and as we will revisit in Sec.~\ref{sec:observables}, there is no contradiction with this intuition because \emph{classical} observables such as the impulse are sensitive only to the combination $_{\rm abs}\sigma^{\text{U}}_P(\omega)-{}_{\rm em}\sigma^{\text{U}}_P (\omega)$. In fact, at the leading order in the $GM\omega$ expansion, one sees
\begin{align}\label{eq:cancellation_of_Hawking_effect_in_cross_section}
    _{\rm abs}\sigma^{\text{U}}_P(\omega)-{}_{\rm em}\sigma^{\text{U}}_P (\omega)=\frac{4\pi}{45\omega^2}(2GM\omega)^6 + \cdots
\end{align}
which coincides with the leading-order absorption cross-section in the Boulware vacuum~\eqref{eq:Boulware_cross_section_aftermatching}.\footnote{The equality $_{\rm abs}\sigma^{\text{U}}_P(\omega)-{}_{\rm em}\sigma^{\text{U}}_P (\omega)={}_{\rm abs}\sigma^{\text{B}}_P(\omega)$ does not hold at higher orders because we have not taken multi-particle processes into account. We will discuss this issue in Sec.~\ref{sec:on-shell_Hawking}.}

\section{Quantum effects in observables}
\label{sec:observables}

In this section, we present concrete applications of our on-shell amplitudes to describe Hawking effects. To relate physical observables and on-shell amplitudes, we will make use of the KMOC formalism~\cite{Kosower:2018adc}, where the initial state of a single semi-classical black hole is given by a wavepacket state
\begin{equation}\label{eq:single-wave-KMOC}
    \ket{\Psi}=\int \dd\Phi(p) \: \phi(p) e^{ib\cdot p} \ket{p},
\end{equation}
with $\phi(p)$ being the wavefunction normalised to be $\int \dd \Phi(p)|\phi(p)|^2=1$ and $b$ is the position of the wavepacket. Among generic choices of the wavepackets, the semi-classical state corresponds to a wavepacket $\phi(p)$ such that any uncertainty associated with its kinematical variables is negligible, and observables are insensitive to details of the wavepackets. In particular, we assume that the result of integration of the wavepackets is to localize the $p$ integral over a given classical momentum of the object $p_{\rm cl}$:
\begin{align}
\label{wavepacket_int}
\int   \dd\Phi(p) |\phi(p)|^2 f(p)=f(p_{\rm cl})+\mathcal{O}(\hbar)
\,.
\end{align} 
See \cite{Kosower:2018adc} for more details. This is our notion of the semi-classical approach based on the amplitudes: in contrast to the   semi-classical approach of Sec. \ref{sec:Quantum_fields_in_Schwarzschild}, where the black hole is a classical background, we first treat the black hole as a quantum state and then choose a specific initial condition so that the quantumness of the black hole state can be negligible. We, on the other hand, keep quantum effects for dynamics described by the amplitudes. In Sec.~\ref{sec:on-shell_Hawking}, we first see that the thermal spectrum of Hawking radiation is reproduced in our formalism. Then, we discuss the implications of quantum effects for binary scattering in Sec.~\ref{sec:Hawking_2body}, especially emphasizing their difference due to the vacuum choice.

\subsection{Hawking radiation from a single black hole}
\label{sec:on-shell_Hawking}
Consider an isolated black hole and an observer on $\mathscr{J}^+$. We consider the problem of counting the number of emitted particles due to Hawking radiation. This problem was already studied in Sec.~\ref{sec:Quantum_fields_in_Schwarzschild}, and the information there has been used to determine the amplitudes in~Sec.~\ref{sec:on-shell}. However, there are two things amiss in the discussions of~Sec.~\ref{sec:on-shell}. The first is the treatment of the black hole; as explained at the beginning of this section, the semi-classical black hole should be given by a wavepacket state~\cite{Kosower:2018adc}. The second is that we have only computed the three-point amplitudes, while the Hawking emission would require higher-point processes as well. Let us discuss these issues in this section.

Choosing \eqref{eq:single-wave-KMOC} as our initial state and setting the rest frame of the wavepacket with $b=0$, the number density of emitted particles is
\begin{align}
    n_{\eta}(k)= \bra{\Psi}\hat{S}^{\dagger} \hat{a}^{\dagger}_{\eta}(k)\hat{a}_{\eta}(k) \hat{S}\ket{\Psi}
    \,.
    \label{eq:number_k}
\end{align}
This number density is different from \eqref{eq:number} (for gravitons) because \eqref{eq:number} is the number density in the spherical basis of gravitons, while \eqref{eq:number_k} is that in the plane-wave basis. The mismatch between the two can be resolved by noting that the creation operator in the spherical basis is given by \cite{Aoude:2023fdm}\footnote{This relation is obtained by requiring the algebra of commutators in a plane wave $[\hat{a}^{\dagger}(k),\hat{a}(k')]=2\omega \hat{\delta}^{3}(\bm{k}-\bm{k}')$ and partial wave basis $[\hat{a}^{\dagger}_{J},\hat{a}_{J'}]=2\omega \hat{\delta}(\omega-\omega')\delta_{JJ'}$ be preserved.}
\begin{equation}
 \hat{a}^{\dagger}_{\eta,J}=\frac{\omega}{2 \pi}\int \dd^{2}\hat{\bm{k}}\: {}_{-\eta}Y_{lm}(\hat{\bm{k}}) \hat{a}^{\dagger}_{\eta}(k)  \, ,
\end{equation}
where ${}_{-\eta}Y_{lm}$ is the spin-weighted spherical harmonics and $\hat{\bm{k}}=\bm{k}/\omega$ with $k=(\omega, \bm{k})$. Alternatively, we can look at the total number of particles
\begin{equation}\label{eq:total-number}
    N_{\rm tot}=
    \begin{dcases}
    \sum_{\eta=\pm 2}\int \dd\Phi(k) \bra{\Psi}\hat{S}^{\dagger} \hat{a}^{\dagger}_{\eta}(k)\hat{a}_{\eta}(k)S\ket{\Psi}
    \qquad&\text{(on-shell)}\,, \\
    \sum_{P={\rm RW,Z}} \int_{J} n^{P}_{{\rm out},J}= \sum_{l=2} \int \frac{\hd \omega}{2\omega}\,2(2l+1)n_{{\rm out},l}(\omega)
    \qquad&\text{(BHPT)}
    \,,
    \end{dcases}
\end{equation}
as it is independent of the basis. Here, $n^{P}_{{\rm out},J}$ is the number density of emitted gravitons computed in the BHPT. It is independent of the magnetic quantum number $m$ and the parity $P$ for the Schwarzschild black hole, yielding the last expression.

The wavepacket initial state (\ref{eq:single-wave-KMOC}) yields
\begin{align}\label{eq:single-wave-KMOC2}
n_{\eta}(k)&=\int \dd\Phi(p,p') \phi(p) \phi^{*}(p')\bra{p'}S^{\dagger}\hat{a}^{\dagger}_{\eta}(k)\hat{a}_{\eta}(k)S\ket{p} 
\,.
\end{align}
Conservation of momentum requires that the incoming and outgoing momenta are the same, so we factor out an overall momentum-conserving delta function
\begin{align}
    \bra{p'}S^{\dagger}\hat{a}^{\dagger}_{\eta}(k)\hat{a}_{\eta}(k)S\ket{p} =\hdelta^{(4)}(p-p')\langle n_{\eta}(k) \rangle_p
    \,.
\end{align}
We perform the on-shell integrals of $p$ and $p'$ by means of \eqref{wavepacket_int} to find
 \begin{align}
     n_{\eta}(k)
     &=\int \dd\Phi(p, p') |\phi(p) |^2 \hdelta^{(4)}(p-p')\langle n_{\eta}(k) \rangle_p
     \nn
     &=\hdelta(E_{\rm cl}-E_{\rm cl})  \frac{\langle n_{\eta}(k) \rangle_{p_{\rm cl}}}{2E_{\rm cl}} 
     \,.
 \end{align}
This quantity is divergent in exactly the same manner as the one we already encountered in Sec.~\ref{sec:Quantum_fields_in_Schwarzschild}. We therefore take the same regularisation procedure of introducing a cutoff in the time integral
\begin{equation}
\hdelta(E_{\rm cl}-E_{\rm cl})=\int^{1/\omega}_{-1/\omega}\dd t= \frac{1}{2\omega}
\,.
\end{equation}
As a result, the number density after the regularisation is
\begin{align}
     n_{\eta}(k)
     &= \frac{\langle n_{\eta}(k) \rangle_{p_{\rm cl}}}{4M \omega} 
     \,.
 \end{align}
From this expression, one sees that the number density is given by the square of the amplitudes with the fixed momentum $p_{\rm cl}=(M,\bm{0})$. This justifies the matching procedure in Sec.~\ref{sec:Quantum_fields_in_Schwarzschild} where the black hole is described as a momentum eigenstate. We, however, note that interference of different momentum states composing the wavepacket is crucial for two-body dynamics~\cite{Kosower:2018adc}, which we will discuss in the next subsection.

We proceed to compute the number density. Using the completeness relation, we find
\begin{align}
    \langle n_{\eta}(k) \rangle_{p_{\rm cl}}
    =\sum_{s,A} \int \dd \mu^2 \dd \Phi(p_X) \rho_{s,A}(\mu^2) (A_{3,A}^{-\eta,s})^* A_{3,A}^{-\eta,s} \hdelta^{(4)}(p_X+k-p_{\rm cl})
    + \cdots
\end{align}
where the ellipsis stands for contributions from multi-particle states. This equation is diagrammatically analogous to an on-shell cut of the physical channel by which the particle decays\footnote{Something similar was argued in \cite{Caron-Huot:2023vxl}.}
\begin{equation}\label{eq:number_density_as_a_cut_of_amplitude}
\langle n_{\eta}(k) \rangle_{p_{\rm cl}}=
\begin{tikzpicture}[baseline=-2]
\begin{feynhand}
  \propag (-1.8, 0) node[left] -- (-0.7,0) ;
  \draw[very thick] (-0.34,0) -- (0.7,0);
  \propag (0.7,0) -- (1.8, 0) node[right]  ;
  \propag [boson] (-0.7,0) .. controls (0,0.9) .. (0.7,0);
  \draw[red, dashed] (0, 0.9) -- (0, -0.3);
  \vertex[grayblob] at (-0.7,0){};
  \vertex[grayblob] at (0.7,0){};
\end{feynhand}
\end{tikzpicture}
+\cdots \,,
\end{equation}
but we do not take the summation over helicities and momenta of the internal graviton yet.
For simplicity, we only compute the decay into the spin-$2$ state, corresponding to the $l=2$ mode emissions
\begin{align}
     \langle n^{s=2}_{\eta}(k) \rangle_{p_{\rm cl}}
     &=80 \pi \varrho^{s=2}_{\eta,-\eta}(M^2-2M\omega)
     \nn
     &=
     \begin{dcases}
         0 &(\text{Boulware vacuum})\,,\, \\
         \frac{20 \pi M}{\omega}\frac{|T|^2e^{-8\pi GM\omega}(1-e^{-8\pi GM\omega})}{\left(1-|R|^2e^{-8\pi GM\omega}\right)^2}
      \qquad &(\text{Unruh vacuum)}
      \,.
     \end{dcases}
\end{align}
The Boulware case vanishes because the discontinuity $\varrho^{s=2}_{\eta,-\eta}(\mu^2)$ is zero for $\mu^2<M^2$. We thus consider the Unruh case only.
The total number of emitted particles from the two-body decay into the spin-2 state is
\begin{align}
    N_{p\to p_X+k}
    &= \sum_{\eta=\pm 2} \int \dd \Phi(k)  \frac{5 \pi }{\omega^2}\frac{|T|^2e^{-8\pi GM\omega}(1-e^{-8\pi GM\omega})}{\left(1-|R|^2e^{-8\pi GM\omega}\right)^2}
    \nn
    &=\int \frac{\hd \omega}{2\omega}\, 2 \times 5\times \frac{|T|^2e^{-8\pi GM\omega}(1-e^{-8\pi GM\omega})}{\left(1-|R|^2e^{-8\pi GM\omega}\right)^2}
    \,.
    \label{N_3pt}
\end{align}
This result should be compared with the result of BHPT, where the total number of emissions for the $l=2$ mode is
\begin{align}
    N_{l=2}=\int \frac{\hd \omega}{2\omega}\, 2\times 5 \times \frac{|T|^2}{e^{8GM\pi \omega}-1}
    \,.
    \label{N_BHPT}
\end{align}
As expected, the results do not agree because only the three-point amplitudes are included in the computation of \eqref{N_3pt}.

\begin{figure}[t]
 \centering
 \includegraphics[width=1\textwidth]{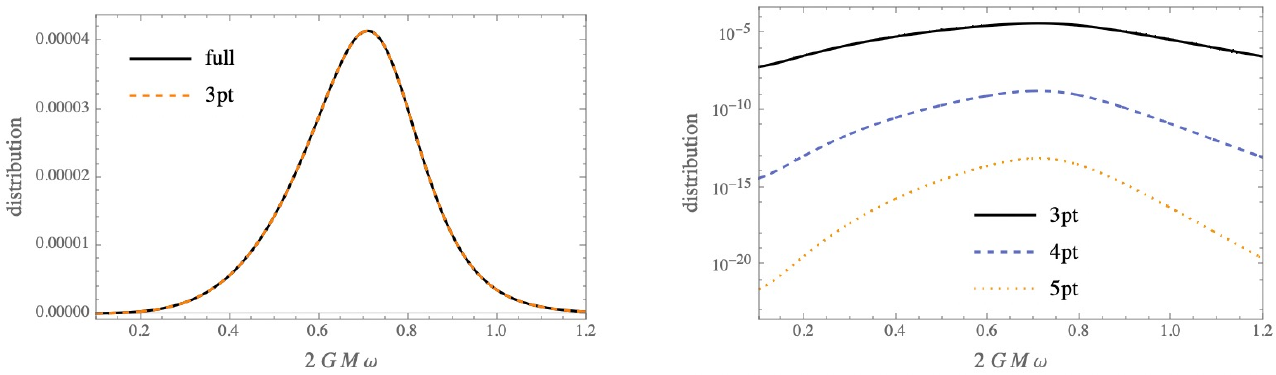}
 \caption{Left: the Hawking thermal distribution and the distribution obtained only via the three-point amplitudes. Right: contributions from $3, 4,$ and $5$ point amplitudes.}
 \label{fig:Hawking-thermal-distribution}
\end{figure}

However, it turns out that \eqref{N_3pt} well approximates the full Hawking spectrum \eqref{N_BHPT}. Let us first look at the low-frequency expansion $GM\omega \ll 1$, where we have
\begin{align}
    |R|^2=1-|T|^2=1 + \mathcal{O}(\omega^6)
    \,.
\end{align}
Hence, we find
\begin{align}
    \frac{|T|^2e^{-8\pi GM\omega}(1-e^{-8\pi GM\omega})}{\left(1-|R|^2e^{-8\pi GM\omega}\right)^2}
    =\frac{|T|^2}{e^{8GM\pi \omega}-1}[1+\mathcal{O}(\omega^6)]
    \,,
\end{align}
showing that the difference between the two is highly suppressed for $GM\omega \ll 1$. On the other hand, in the high-frequency region $GM\omega \gg 1$, the reflectivity $|R|^2$ decays exponentially, giving
\begin{align}
    \frac{|T|^2}{e^{8GM\pi \omega}-1} &\approx |T|^2 e^{-8GM\pi \omega}
    \,, \\
    \frac{|T|^2e^{-8\pi GM\omega}(1-e^{-8\pi GM\omega})}{\left(1-|R|^2e^{-8\pi GM\omega}\right)^2}
    &\approx |T|^2 e^{-8GM\pi \omega}\,.
\end{align}
The intermediate region $GM\omega =\mathcal{O}(1)$ requires a numerical analysis of computing the greybody factor. In the left panel of Fig.~\ref{fig:Hawking-thermal-distribution}, we show the distributions of the number density obtained by summing over all emission processes (the Hawking spectrum) and by the three-point process only.\footnote{Note that the three-point amplitudes have only one graviton emission, but this does not mean that the number of emitted particles is one. The number density has been computed by summing over all three-point processes.} We also show the contributions to the number density from four-point (two-graviton) amplitudes and five-point (three-graviton) amplitudes in the right panel. Here, we employed numerically computed transmission coefficients using a seventh‑order WKB approximation~\cite{Konoplya:2019hlu}. Our analysis shows that the distribution computed by the three-point amplitudes exhibits excellent agreement with the full Hawking spectrum across the full frequency range.

We can also see the suppression of multi-particle processes in classical observables. At the end of Sec.~\ref{sec:matching_Unruh}, we mentioned that classical observables are only sensitive to the difference between the absorption and emission processes
\begin{align}\label{eq:effective_cross_section_for_B_and_U}
    {}_{\rm abs}\sigma^{\rm B}_P(\omega)\,, \qquad {}_{\rm abs}\sigma^{\rm U}_P(\omega)-{}_{\rm em}\sigma^{\rm U}_P(\omega)
    \,.
\end{align}
We numerically compute these quantities by only including the three-point amplitudes in Fig.~\ref{fig:effective_cross_sections_Boulware_and_Unruh}, showing their agreement. All in all, we can confirm that  three-point amplitudes are enough to recover the predictions of BHPT with a good approximation.

\begin{figure}[t]
 \centering
 \includegraphics[width=1\textwidth]{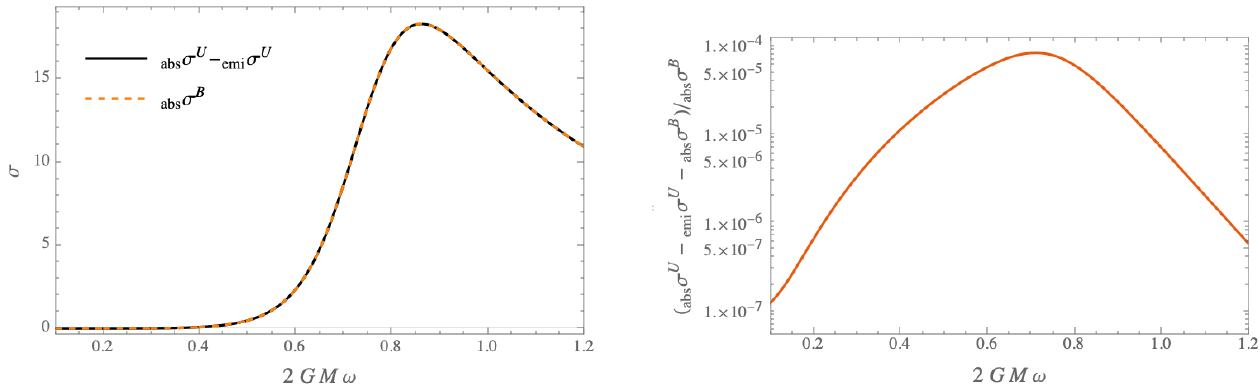}
 \caption{Left: the comparison of the cross-sections in Eq.~\eqref{eq:effective_cross_section_for_B_and_U}. Right: relative error of \eqref{eq:effective_cross_section_for_B_and_U}.}
 \label{fig:effective_cross_sections_Boulware_and_Unruh}
\end{figure}
 
\subsection{Quantum effects in two-body problems}
\label{sec:Hawking_2body}

We now consider a binary system of a black hole (particle $1$, with initial mass $M_1$ and incoming momentum $p_{1,{\rm cl}}^\mu=M_1u_1^\mu$) and a generic compact body (particle $2$, with mass $M_2$ and outgoing momentum $p_{2,{\rm cl}}^\mu = M_2u_2^\mu$). In this framework, we evaluate the Hawking effect on the induced mass shift by employing the absorption and emission amplitudes computed in the previous section. 

Let us consider wave-packet states of two particles
\begin{align}
    \ket{\Psi}=\int \dd \Phi(p_1,p_2)\phi(p_1)\phi(p_2)e^{i(b_1\cdot p_1+b_2\cdot p_2)}\ket{p_1;p_2}
\end{align}
separated by the impact parameter $b:=b_1-b_2$. The problem involves three distinct length scales: the Compton wavelength $\ell_{c}$, which characterizes the intrinsic quantum size of the particle; the wavepacket spread $\ell_{w}$, which reflects the wave nature of the object; and the separation scale between the two objects, $\ell_{s}\sim\sqrt{-b^{2}}$. The (semi-)classical two-body problem calls for the so-called \textit{Goldilocks} condition~\cite{Kosower:2018adc}
\begin{equation}
    \ell_c \ll \ell_w \ll \ell_s\,,
\end{equation}
under which both the quantum and the wave natures of objects can be negligible.

Exploiting this hierarchy of scales, we evaluate the mass shift of particle~$1$ induced by the interaction with the companion object under the double expansion of $G$ and $\hbar$.\footnote{This should be distinguished from the spontaneous mass shift due to the Hawking radiation. At the diagrammatic level, the spontaneous emission comes from diagrams where particles 1 and 2 are disconnected. We only compute the induced one corresponding to connected diagrams. } The squared mass is computed by the expectation value of $\hat{P}_1^2$ with $\hat{P}_1^\mu$ being the momentum operator for particle $1$.
When the mass-shift $\Delta M_1$ is small compared to the initial mass $M_1$, it can be approximated as $\frac{\Delta M^2_1}{2M_1}$, so it is given by~\cite{Jones:2023ugm, Bautista:2024emt}
\begin{align}
    \ev{\Delta M_1} = \bra{\Psi}\hat{S}^\dagger \frac{\hat{P}_1^2-M_1^2}{2 M_1} \hat{S} \ket{\Psi}.  
\end{align}
It amounts to the longitudinal contribution of the impulse for $\hat{P}_1^\mu$. At the leading order in the gravitational coupling, the mass shift is computed by the transition into the mass-changing particle 1 while preserving the mass of particle 2:
\begin{equation}\label{eq:mass_shift_KMOC_formula}
\begin{aligned}
\ev{\Delta M_1}_{\rm LO} &= \sum_{s, A} \int \dd\Phi(p_2,p_X) \dd \mu^2 \rho_{s, A}(\mu^2) \frac{(\mu^2 - M_1^2)}{2M_1}  \bra{\Psi} \hat{S}^{\dagger} \ket{p_X,s,A; p_2} \bra{p_X,s,A; p_2} \hat{S} \ket{\Psi}\\
&\simeq \sum_{s,A} \int \dd\mu^2 \rho_{s,A}(\mu^2)\frac{\mu^2-M_1^2}{2M_1}\int \dd\Phi(p_X, p_2'')\hat{\dd}^4q\frac{\hat{\delta}(u_1\cdot q)\hat{\delta}(u_2\cdot q)}{4 M_1M_2}e^{-ib\cdot q}\\
&\times 
\hat{\delta}^{(4)}(p_1+p_2-p_X-p_2'')\mathcal{A}_4^*\left(p_1+q,p_2-q\to p_X,p_2''\right)\mathcal{A}_4\left(p_1,p_2\rightarrow p_X,p_2''\right)~.
\end{aligned}
\end{equation}
In the second line, the symbol $\simeq$ denotes the semiclassical limit, wherein the external momenta are identified with their classical values, $p_{1}=M_{1}u_{1}$ and $p_{2}=M_{2}u_{2}$ (with omitted the subscript ``cl''). The four-point amplitudes are computed at the tree level at the leading order in $G$. Eq.~\eqref{eq:mass_shift_KMOC_formula} then corresponds to the following cut of a $1$-loop amplitude
\begin{align}
\begin{tikzpicture}[baseline=-2]
\begin{feynhand}
\propag (2, -0.7) node[right] {$p_2$} -- (-2, -0.7) node[left] {$p_2'=p_2-q$};
\propag (2,0.7) node[right] {$p_1$} -- (0.8,0.7);
\propag (-2,0.7) node[left] {$p_1'=p_1+q$}-- (-0.8,0.7);
\draw[very thick] (-0.8,0.7) -- (0.8, 0.7) ;
\propag [boson]  (-0.8, -0.7) -- (-0.8, 0.7);
\propag [boson]  (0.8, -0.7) -- (0.8, 0.7);
\vertex[grayblob] at (-0.8,0.7){};
\vertex[grayblob] at (0.8,0.7){};
\draw[dashed] (0, 0.9) node [above] {$p_X$} -- (0, -0.9) node [below] {$p_2''$};
\end{feynhand}
\end{tikzpicture}
\label{mass-shift_diagram}
\,.
\end{align}
At the same time, it is well established that the contribution at the leading order in $\hbar$ arises from singular parts of such amplitudes. Within the framework of the generalized unitarity method~\cite{Bern:1994cg,Bern:1995db,Bern:1994zx,Bern:1997sc,Britto:2004nc}, this singular contribution is extracted from generalised unitarity cuts, thereby demonstrating that the leading $\hbar$ piece is completely determined by the maximal cut~\cite{Jones:2023ugm,Bautista:2024emt}, consisting of four on-shell three-point amplitudes, as diagrammatically shown in 
\begin{align}
\begin{tikzpicture}[baseline=-2]
\begin{feynhand}
\propag (2, -1) node[right] {$p_2$} -- (-2, -1) node[left] {$p_2'=p_2-q$};
\propag (2,1) node[right] {$p_1$} -- (0.8,1);
\propag (-2,1) node[left] {$p_1'=p_1+q$}-- (-0.8,1);
\draw[very thick] (-0.8,1) -- (0.8, 1) ;
\propag [boson]  (-0.8, -1) -- (-0.8, 1);
\propag [boson]  (0.8, -1) -- (0.8, 1);
\vertex[grayblob] at (-0.8,1){};
\vertex[grayblob] at (0.8,1){};
\draw[dashed] (0, 1.4) node[above] {$p_X$} -- (0, -1.4) node[below] {$p_2''$};
\draw[dashed] (-1.2, 0) node[left] {$\ell'=\ell-q$} -- (1.2, 0) node[right] {$\ell$};
\node at (-0.8, 0.3) {$+~~-$} ;
\node at (-0.8, -0.3) {$-~~+$};
\node at (0.8, 0.3) {$+~~-$} ;
\node at (0.8, -0.3) {$-~~+$};
\end{feynhand}
\end{tikzpicture}
\,.
\label{maximal_cut}
\end{align}
Then, the computation of Eq.~\eqref{eq:mass_shift_KMOC_formula} is reduced to glueing four three-point amplitudes.

The mass-changing three-point amplitudes are introduced in Eq.~\eqref{eq:3-point-EFT}. On the other hand, the three-point equal-mass amplitudes minimally-coupled to a graviton with momentum $\ell^\mu$ are given by~\cite{Arkani-Hamed:2017jhn}:
\begin{equation}
\label{eq:3pt_amp_minimal_coupling}
\mathcal{A}^{+}_{3}
:=-\frac{\kappa}{2} \frac{\left[\ell\left|p_2\right| \xi\right\rangle^2}{\langle \ell \xi\rangle^2}\,, \quad \mathcal{A}^{-}_{3}
:= -\frac{\kappa }{2} \frac{\left.\langle \ell| p_2 | \xi\right]^2}{[\ell \xi]^2}\,,
\end{equation}
where $\kappa = \sqrt{32\pi G}$ is the gravitational coupling and $\xi$ is a reference massless spinor related to the gauge freedom.
Notice that the mass change contribution can be combined and recast into the form of $\text{Disc}_X\mathcal{A}_4^{\eta',\eta}$. Hence, we find
\begin{align}\label{eq:mass_shift_in_terms_of_calB}
   \ev{\Delta M_1}_{\rm LO} &=2M_1 \int \dd w w \int \hat{\dd}^4q \frac{\hat{\delta}(u_1\cdot q)\hat{\delta}(u_2\cdot q)}{4M_1M_2}e^{-ib\cdot q} \mathcal{B}(w,q)
\end{align}
where
\begin{equation}
\begin{aligned}
    \mathcal{B} (w,q) := 
    &  \sum_{\eta, \eta'}     
    \mathcal{N}^{\eta,\eta'}
     \int \hat{\dd}^4 \ell \frac{\hat{\delta}(u_1\cdot \ell-w)\hat{\delta}(u_2\cdot \ell)}{4M_1M_2 \ell^2(\ell-q)^2} 
\end{aligned}
\end{equation}
and
\begin{align}
    \mathcal{N}^{\eta,\eta'}:=\sum_{\eta, \eta'}     
     {\rm Cut}\left[ \frac{1}{2\pi i} \text{Disc}_X\mathcal{A}_4^{\eta, \eta'} (w)\mathcal{A}_3^{\eta *} (p_2-q \to p_2'', \ell-q)\mathcal{A}_3^{-\eta'}(p_2 \rightarrow p_2'',\ell)\right]
     \,.
\end{align}
Here, $w$ is the mass-changing parameter of particle $1$ defined by
\begin{equation}\label{masschange}
    \begin{split}
        w:=\frac{\mu^2-M_1^2}{2 M_1}\,.
    \end{split}
\end{equation}
The symbol ${\rm Cut}[\cdots ]$ denotes that the momenta of the expression inside are fixed by the cut conditions as specified by \eqref{maximal_cut}.

Because the graviton carries the polarization, $\mathcal{B} (w,q)$ has contributions from four distinct structures depending on their helicities $\mathcal{N}^{\eta,\eta'}$.
By using \eqref{eq:disc_eq_varrho_P}, \eqref{eq:disc-struc_preserve} and \eqref{eq:disc-struc_violate} for the discontinuity and substituting \eqref{eq:3pt_amp_minimal_coupling} for equal mass three-point amplitudes, we have
\begin{align}
\mathcal{N}^{+,-}+\mathcal{N}^{-,+} 
&=\frac{5 \kappa^2M_2^4\varrho_{+,-}}{8 w^4}[\ell'|u_1\ket{\ell}^{2} \bra{\ell'}u_2 u_1 u_2 |\ell]^2  + (+\leftrightarrow -) \nonumber \\
&=
\frac{5\kappa^2M_2^4 \left(\varrho_{+,-} + \varrho_{-,+}\right)}{8 w^4}
(16 w^4-8w^2(1-4\gamma^2)|q|^2+(1-8\gamma^2 + 8\gamma^4)|q|^4) \nonumber 
\\&\label{eqeps}\,\,\,\,\,\,\,\,+\frac{5 \kappa^2M_2^4(\varrho_{+,-}-\varrho_{-,+})}{w^4} \, i \gamma \left(4 w^2-\left(1-2 \gamma ^2\right) |q|^2\right)  \varepsilon(q,\ell, u_1, u_2) 
\,, \\
\mathcal{N}^{+,+}+\mathcal{N}^{-,-} 
&=\frac{5\kappa^2 M_2^4 \varrho_{+,+}}{8 w^4}[\ell'\ell]^2 \langle \ell'\ell\rangle^2 + (+\leftrightarrow -)
=
\frac{\kappa^2 M_2^4 \left(\varrho_{+,+}
+\varrho_{-,-}
\right)}{64w^4} |q|^4\, ,
\end{align}
where $\gamma=u_1\cdot u_2$ denotes the Lorentz factor, $|q|^2=-q^2>0$ and $\varepsilon (q,\ell, u_1, u_2) = \varepsilon^{\mu\nu \rho \sigma} q_\mu \ell_\nu u_{1,\rho} u_{2,\sigma}$ is the Levi-Civita tensor contracted with the four momenta in the parentheses. As shown in Eq.~\eqref{eq:Unruh-rho_helicity_perserving} and Eq.~\eqref{eq:Unruh-rho_helicity_violating}, the spherical symmetry ensures $\varrho_{+,-}=\varrho_{-,+}$ and $\varrho_{+,+}=\varrho_{-,-}$, guaranteeing the absence of the imaginary part of the helicity-preserving contribution, the third line of \eqref{eqeps}. In addition, the helicity-violating spectral contribution vanishes $\mathcal{N}^{+,+/-,-}$, as ensured by electric–magnetic duality. In the end, we are left with
\begin{equation}
\begin{aligned}
    \mathcal{B} (w,q)  
    =
    \frac{5 \kappa^2 M_2^3 \varrho_{+,-} (w)}{16 M_1  w^4 \sqrt{\gamma^2-1}} 
    &(16 w^4-8w^2(1-4\gamma^2)|q|^2+(1-8\gamma^2 + 8\gamma^4)|q|^4) \\
    &\times 
    \int\hat{\dd}^2 \ell_{\perp} \frac{ 1}{\left[\ell_\perp^2 + \frac{w^2}{\gamma^2-1}\right] \left[(\ell-q)_\perp^2 +\frac{w^2}{\gamma^2-1}\right]}\,,
\end{aligned} 
\end{equation}
after performing the two components of the $\ell$-integral by means of the delta functions.
The expression is evaluated on the on-shell phase space integral support in Eq.~\eqref{eq:mass_shift_in_terms_of_calB}. Upon performing the remaining $\ell$- and $q$-integral,\footnote{See also Appendix E of \cite{Jones:2023ugm}.}  one obtains the mass shift as follows
\begin{align}
    \ev{\Delta M_1}_{\rm LO} =2 M_1 \int_{-\infty}^\infty \dd w w  \varrho_{+-}(w) \mathcal{C} (w,b),
\end{align}
where $\mathcal{C}(w,q)$ is the universal building block for absorption and emission processes

\begin{align}
    \mathcal{C} (w,b) 
    =\frac{5\kappa^2 M_2^2}{256 M_1^2 w^4 (\gamma^2-1) \pi^2} 
    \left[16 w^4+8w^2(1-4\gamma^2)\nabla_b^2+(1-8\gamma^2 + 8\gamma^4)\nabla_b^4
    \right] K_0^2.
\end{align}
Here, $ K_0:=  K_0( |b| \sqrt{w^2/(\gamma^2-1)}\,)$ is the modified Bessel function of the second kind. Above, we also introduced the projected derivatives as in~\cite{Menezes:2022tcs}
\begin{equation}
\begin{split}\label{projct}
\nabla_b^\mu:= {\Pi^{\mu}}_{\nu} \frac{\partial}{\partial b^\nu},\,\,\,\,\,\,\,{\Pi^{\mu}}_{\nu} :=\frac{\varepsilon^{\mu\alpha}(u_1, u_2)\varepsilon_{\alpha\nu}(u_1, u_2)}{\gamma^2-1},
\end{split}
\end{equation}
 with $\varepsilon^{\mu\alpha}(u_1, u_2):= \varepsilon^{\mu\alpha\rho\gamma}u_{1,\rho} u_{2,\, \gamma}$ and $\Pi^2=\Pi$ is a projector orthogonal to the black hole velocities first introduced in \cite{Vines:2017hyw}.\par

It is important to notice that the universal part $\mathcal{C}(w,b)$ is an even function of $w$. The induced mass shift can thus be written as
 \begin{align}
 \bigl\langle\Delta M_1\bigr\rangle_{\rm LO}
&=2M_1\int_{0}^{\infty}\dd w\;w\,\varrho_{+,-}(w)\,\mathcal{C}(w,b)
+2M_1\int_{-\infty}^{0}\dd w\;w\,\varrho_{+,-}(w)\,\mathcal{C}(w,b)
\nn
&=2M_1\int_{0}^{\infty}\dd w\;w\,
  \bigl[\varrho_{+,-}(w)-\varrho_{+,-}(-w)\bigr]
  \,\mathcal{C}(w,b)
  \,.
 \end{align}
 Therefore, the mean value of the mass shift is only sensitive to the combination $\varrho_{+,-}(w)-\varrho_{+,-}(-w)$, and thus ${}_{\rm abs}\sigma(\omega)-{}_{\rm em}\sigma(\omega)$. The situation is different for \emph{quantum} observables, such as the variance
 \begin{align}
     \ev{\sigma_1}
    &:= \bra{\Psi} \hat{S}^\dagger \left(\frac{\hat{P}_1^2-M_1^2}{2 M_1}\right)^2 S\ket{\Psi} - \left(\bra{\Psi} \hat{S}^\dagger \frac{\hat{P}_1^2-M_1^2}{2 M_1}\hat{S} \ket{\Psi}\right)^2
    \,.
\end{align}
At leading order in $G$, the second term is negligible because it is a square of $\varrho_{\eta,\eta'}$ and suppressed by a higher-order in $G$. Hence, the variance at leading order takes the form
\begin{align}
    \ev{\sigma_1}_{\rm LO}&= \bra{\Psi}\hat{S}^\dagger \left(\frac{\hat{P}_1^2-M_1^2}{2 M_1}\right)^2
    \hat{S}\ket{\Psi}
    \nn
    &=2M_1\int_{-\infty}^{\infty}\dd w\;w^2\,
  \varrho_{+-}(w) \mathcal{C}(w,b)
  \nn
  &=2M_1\int_{0}^{\infty}\dd w\;w^2\,
  \bigl[\varrho_{+,-}(w)+\varrho_{+,-}(-w)\bigr]
  \,\mathcal{C}(w,b)
 \end{align}
 which depends on the sum $\varrho_{+,-}(w)+\varrho_{+,-}(-w)$, instead of the difference.

Let's compute these observables concretely. For the convenience of readers, we recapitulate the soft expansion of $\varrho_{+,-}$ up to $\mathcal{O}(w^5)$:
\begin{align}
    \varrho^{\rm B}_{+,-}(w)=
     \begin{dcases}
 \frac{64}{225} G^6M_1^7  w^5
 &(w\to 0^+)\\\
  0& (w\to0^-)
\end{dcases}
\qquad \text{(Boulware vacuum)}
\,,
\end{align}
and
\begin{equation}\label{eq:spectral_density_with_Hawking_effect}
\varrho^{\text{U}}_{+,-} (w)
 =
 \begin{dcases}
 \frac{8}{225\pi} G^5 M_1^6 w^4 + \frac{16}{75} G^6M_1^7  w^5 &(w\to 0^+)\\
  \frac{8}{225\pi} G^5 M_1^6 w^4 -\frac{16}{225} G^6 M_1^7  w^5  & (w\to0^-) 
\end{dcases}
\qquad \text{(Unruh vacuum)}
\,.
\end{equation}
While the infinite integrals would diverge by substituting the low-frequency expansions, we can regularize the integrals by following~\cite{Jones:2023ugm} (which is similar to the dimensional regularization). The observables are then computed as, after restoring Planck constant $\hbar$, 
\begin{align}
\label{mean_value}
    \bigl\langle\Delta M_1\bigr\rangle_{\mathrm{LO}}^{\rm B}
&=\bigl\langle\Delta M_1\bigr\rangle_{\mathrm{LO}}^{\rm U}=\frac{5\hbar^0 G^7 \pi M_1^6 M_2^2 \sqrt{\gamma^2-1} (21\gamma^4 - 14\gamma^2 +1)}{16|b|^7}\,,
\end{align}
and
\begin{align}
    \ev{\sigma_1}^{\rm B}_{\text{LO}} 
    &= \frac{1024 \hbar^1 G^7 M_1^6 M_2^2 (\gamma^2-1) (80\gamma^4-48\gamma^2 +3)}{525\pi |b|^8}
    \,,
    \label{variance_B}
    \\
    \ev{\sigma_1}^{\rm U}_{\text{LO}} 
    &= \frac{5\hbar^1 G^6 M_1^5 M_2^2 \sqrt{\gamma^2-1} (21\gamma^4 - 14\gamma^2 +1)}{64 |b|^7}
    \,.
    \label{variance_U}
\end{align}
 The mean value is independent of both $\hbar$ and the choice of vacuum (see also~\cite{Goldberger:2020wbx,Jones:2023ugm}). This confirms that the quantum nature of black holes does not affect the classical observables. On the other hand, the variance is of the order of $\hbar$ and depends on the vacuum choice. Note that the variance of the Boulware vacuum is suppressed by $GM_1/|b|$ in comparison to the Unruh case. This suppression is consistent with the intuition that black holes in the Boulware vacuum are ``less-quantum'' than those in the Unruh vacuum in the sense of no thermal radiation.

\section{Conclusions}\label{sec:conclusions}

Hawking’s discovery that black holes emit thermal radiation has profoundly altered our understanding of gravity and quantum field theory. Black holes are not immortal, but evolve through the interplay of quantum effects at the horizon, gradually losing their mass. Yet, despite decades of work, the research on the Hawking effect is still an active field and remains beyond reach in complex situations such as two-body problems and effects on emissions of gravitational waves. Since any departure from the classical black hole picture would necessarily imply deviations from perfect absorption, such modifications are expected to play a role in future experiments. 
Given the recent advances in relating classical observables such as the impulse or waveforms to on-shell scattering amplitudes, one may ask whether Hawking effects can also be cast into such a modern framework. At the classical level, the great advantages of using on-shell amplitudes to describe classical gravity are that they bypass the need for specific gauge choices or fields, allow the use of modern techniques from particle physics, and thereby highlight the simplicity of calculations. It could be useful to apply these developments to study the Hawking effects and to gain new insights into the quantum nature of black holes.

We have shown that such an answer is affirmative. In Section~\ref{sec:Quantum_fields_in_Schwarzschild}, we started to revisit the $S$-matrices in a black hole spacetime discussed in~\cite{Frolov:1998wf}. The structure of $S$-matrices depends on the vacuum choice, representing distinct phenomena: no radiation is detected for black holes in the Boulware vacuum, and the well-known Hawking's thermal distribution of emitted particles is obtained for those in the Unruh vacuum. Building on this knowledge, in Section \ref{sec:on-shell}, we constructed an alternative description in which the black hole is treated as a “particle” and only on-shell quantities on the boundary Minkowski spacetime are considered. Both classical absorption and Hawking effects can be described by excitation/decay processes where a particle gains/loses its mass by absorbing/emitting quanta. Interestingly, since we know the explicit $S$-matrix in a ``fundamental'' theory, we can also gain insight into the dynamics of the invisible sector, which appears only as internal dynamics of a composite particle in the ``effective'' on-shell description (see Figures~\ref{fig:composite_particle} and~\ref{fig:xstate_Unruh}). After determining this building block, in the spirit of the amplitudes program, we computed physical observables in Section~\ref{sec:observables}. It turned out that Hawking's thermal distribution is well understood by a collection of three-point processes. Quantum effects on a binary observable were computed by applying the on-shell techniques. They demonstrate that the modern on-shell approach can indeed help us to develop our understanding of black holes.

Quantum effects on binaries have interesting implications. A key advantage of our framework, along the lines of effective theories, is that the description does not necessarily rely on specific ``fundamental'' theories --- the vacuum choice in our context --- until the end of calculations. We have discussed the mass shift, especially its variance computed here for the first time, of a black hole induced by the binary motion.\footnote{It has been argued by some of the present authors that its vanishing is closely tied to the leading $\hbar$ behaviour of on-shell amplitudes, while its strict disappearance in the classical limit carries additional implications for the underlying quantum theory~\cite{Cristofoli:2021jas}.} First, the mean value is given by the difference between the absorption and emission probabilities. It carries no $\hbar$ scaling and is the same in both Boulware and Unruh vacua, consistently with the fact that the quantum effects should not modify classical observables~\cite{Goldberger:2020wbx, Goldberger:2020geb}. On the other hand, the absorption and emission contribute to the variance with the same sign, leading to the vacuum dependence. The variance is non-zero at $\mathcal{O}(\hbar)$ and, in the large-distance expansion, it is one order of magnitude larger in the Unruh vacuum than the Boulware vacuum, as might be anticipated by the existence of Hawking radiation.

While we focused on the Schwarzschild black hole in this work, our approach is intrinsically universal and general. The on-shell description should be applicable to other objects, such as ordinary macroscopic objects like stars, black holes beyond general relativity, black hole mimickers, and even microscopic black holes in quantum gravity. Their internal differences are encapsulated in the discontinuities, serving as a universal framework to investigate their absorption and emission dynamics at both classical and quantum levels. For instance, a natural extension is to consider the Kerr geometry, which exhibits additional phenomena such as superradiance that can be similarly understood by absorption and emission effects~\cite{Frolov:1998wf, Endlich:2016jgc}. 
Furthermore, having established building blocks capturing the Hawking effect, we can employ them to compute the observables associated with more intricate scattering processes arising from the non-linearity of gravity. A particularly important observable, directly relevant for gravitational-wave detection, is the waveform~\cite{Cristofoli:2021vyo}, which is provided by five-point amplitudes. It is interesting that, as done in this article, much of the physics can be understood in terms of the analytical properties of the amplitudes, as outlined in \cite{Elkhidir:2023dco}.
Computing this is also worthwhile, since the presence of the massless leg renders it non-trivial whether the quantum correction to the waveform survives. We leave these exciting new directions for future work.

\acknowledgments

We are grateful to Donal O'Connell, Asaad Elkhidir, Rafael Aoude and Yu-Tin Huang for their comments and feedback on a draft of our manuscript. We also thank Dogan Akpinar, Toshifumi Noumi, Kazumasa Okabayashi, and Naritaka Oshita for useful discussions. The work of K.A. was supported by JSPS KAKENHI Grant Nos.~JP24K17046 and JP24KF0153. A.C. was supported by JSPS KAKENHI Grant No. JP24KF0153. The work of H.~J. is supported by the JSPS KAKENHI Grant No.~JP24KJ0902, and World Premier International Research Center Initiative (WPI Initiative), MEXT, Japan. M.S. has been supported by the European Research Council under Advanced Investigator grant [ERC–AdG–885414]. K.Y. was supported by JST SPRING, Grant No.~JPMJSP2108 and by a research
encouragement grant from the Iwanami Fujukai. K.A. and A.C. are particularly grateful to the University of Tokyo, Komaba, and their group for their warm hospitality multiple times where part of this work was developed.
H.J., M.S., and K.Y.~also wish to thank the Yukawa Institute for Theoretical Physics for its hospitality and the workshop ``Gravity 2025: New Horizon of Black Hole Physics'' (YITP-W-24-20), where this research was initiated and conducted.

\appendix

\section{Thermal state as squeezed state}
\label{sec:Thermal_Phenomenon_as_a_squeezed state}

As we have seen in Sec.~\ref{sec:Quantum_fields_in_Schwarzschild}, in order to describe the effects of Hawking radiation, it is necessary to introduce an additional set of modes in the invisible sector, the ``dn" modes, in addition to the set of modes defined in the Boulware vacuum. The introduction of such extra degrees of freedom to account for thermal effects is, in fact, a rather general feature. In the following appendix, we provide an explanation of this procedure both in quantum mechanics and quantum field theory.

\subsection{Thermal state in quantum mechanics}
\label{subsec:Thermal_effets_in_Quantum_Mechanics}

Thermal effects in quantum mechanics are discussed in detail in Refs.~\cite{Umezawa:1982nv,Umezawa:1993yq,gerry2023introductory}. To illustrate the basic idea, we consider the simplest case: a harmonic oscillator, denoted by $(a, a^\dagger)$ (in Appendix~\ref{sec:Thermal_Phenomenon_as_a_squeezed state}, the operators are denoted without a hat). The thermal degree of freedom is introduced by augmenting the system with an additional, commuting harmonic oscillator $( \Tilde{a}, \Tilde{a}^\dagger )$:

\begin{equation}
    \comm{a}{a^\dagger} = \comm{\Tilde{a}}{\Tilde{a}^\dagger} = 1\,, \qquad
    \comm{a}{\Tilde{a}} = \comm{a}{\Tilde{a}^\dagger} = 0\,. 
\end{equation}
We introduce the unitary operator

\begin{equation}
    U_{\text{B}} (\theta) = \exp \qty[i\theta G_{\text{B}}]
\end{equation}
where $G_{\text{B}} = i\qty[a\Tilde{a} - \Tilde{a}^\dagger a^\dagger]$ is the generator and $\theta \in \mathbb{R}$. Then the following transformation
\begin{equation}
    \alpha (\theta) = U_{\text{B}} (\theta) a U^{-1}_{\text{B}} (\theta)\,,\qquad
    \Tilde{\alpha} (\theta) = U_{\text{B}} (\theta) \Tilde{a} U^{-1}_{\text{B}} (\theta)\,,
\end{equation}
defines new harmonic oscillators~$\alpha (\theta), \Tilde{\alpha} (\theta)$: 

\begin{equation}
    \alpha (\theta) = a\cosh\theta  - \Tilde{a}^\dagger \sinh \theta\,, \qquad
    \Tilde{\alpha} (\theta) = \Tilde{a} \cosh \theta - a^\dagger \sinh \theta \,. 
\end{equation}
These are called the thermal Bogoliubov transformation to two-mode squeezed states. \par

At this point, we have two types of vacua. One is defined by the original modes~$(a, \Tilde{a})$ and another is defined by the squeezed modes~$(\alpha (\theta), \Tilde{\alpha} (\theta))$. We denote these as $\ket{0}$ and $\ket{0 (\theta)}$, respectively:

\begin{equation}
a\ket{0} = \Tilde{a}\ket{0} = 0\,,\qquad
 \alpha (\theta) \ket{0(\theta)} = \Tilde{\alpha} \ket{0 (\theta)} = 0\,.
\end{equation}
Then we have the following relation 

\begin{equation}
    \ket{0(\theta)} = U_{\text{B}} (\theta) \ket{0}=\exp\qty[-\ln \cosh \theta] \exp \qty[a^\dagger \Tilde{a}^\dagger \tanh \theta] \ket{0}\,,
\end{equation}
which shows that the $(a,\Tilde{a})$-pairs (also called thermal pairs) are condensed in the vacuum~$\ket{0 (\theta)}$. \par

There are two comments. First, the above Bogoliubov transformation generates the thermal noise in the pure states: 

\begin{equation}
    \ev{(\Delta q)^2}\ev{(\Delta p)^2} \geq \frac{\hbar^2}{4} + \ev{\Delta q \Delta \Tilde{q}} \ev{\Delta p \Delta \Tilde{p}} = \frac{\hbar^2}{4}(1+\sinh^2 2\theta)
\end{equation}
where $q, p$ and $\Tilde{q}, \Tilde{p}$ are constructed from $a, a^\dagger$ and $\Tilde{a}, \Tilde{a}^\dagger$ in a standard way.
Second, the expectation value by the pure state in the doubled Hilbert space~$\mathcal{H} (a, \Tilde{a})$ is equivalent to the statistical average of the mixed state expectation value in the original Hilbert space~$\mathcal{H} (a)$. 

\subsection{Thermal state in quantum field theory}
\label{subsec:Thermal_effets_in_Quantum_Field_Theory}

To address macroscopic ordered states in thermal physics—which inherently require an infinite number of degrees of freedom—we employ the framework of quantum field theory. As discussed in Section~\ref{subsec:Thermal_effets_in_Quantum_Mechanics}, we introduce a doubling of the degrees of freedom to incorporate thermal effects. This formalism, when extended to quantum field theory, is known as Thermo-Field Dynamics (TFD). \par

We prepare two commutable harmonic oscillators:

\begin{equation}
    \comm{a_{\bm{k}}}{a^\dagger_{\bm{l}}} = \comm{\Tilde{a}_{\bm{k}}}{\Tilde{a}^\dagger_{\bm{l}}} = \delta^{(3)} (\bm{k}-\bm{l})\,,\qquad
    \comm{a_{\bm{k}}}{\Tilde{a}_{\bm{l}}} = \comm{a_{\bm{k}}}{\Tilde{a}^\dagger_{\bm{l}}} = 0\,. 
\end{equation}
We define the creation and annihilation operators $(\xi_{\bm{k}} (\theta), \Tilde{\xi}_{\bm{k}} (\theta))$ by the thermal Bogoliubov transformation.

\begin{equation}
\begin{aligned}
    \xi_{\bm{k}} (\theta) &=U_{\text{B}} (\theta) a_{\bm{k}} U^{-1}_{\text{B}}(\theta) =  a_{\bm{k}} \cosh \theta_{\bm{k}} -\Tilde{a}^\dagger_{\bm{k}} \sinh \theta_{\bm{k}}\,,\\
    \Tilde{\xi}_{\bm{k}} (\theta) 
    &=U_{\text{B}} (\theta) \Tilde{a}_{\bm{k}} U^{-1}_{\text{B}}(\theta) = \Tilde{a}_{\bm{k}} \cosh \theta_{\bm{k}} -a^\dagger_{\bm{k}} \sinh \theta_{\bm{k}}\,.
\end{aligned}
\end{equation}

The unitary operator which induces these transformations is 

\begin{equation}
    U_{\text{B}} (\theta) = \exp\qty[i G_{\text{B}} (\theta)]
\end{equation}
where 

\begin{equation}
    G_{\text{B}} = i \int \dd^3 {\bm k} \,\theta_{\bm{k}} \qty[a_{\bm{k}}\Tilde{a}_{\bm{k}} - \Tilde{a}^\dagger_{\bm{k}} a^\dagger_{\bm{k}}]\,.
\end{equation}
By following the same steps as in Appendix~\ref{subsec:Thermal_effets_in_Quantum_Mechanics}, we define the two types of vacua:

\begin{equation}
    a_{\bm{k}}\ket{0} 
    = \Tilde{a}_{\bm{k}}\ket{0} = 0\,,\qquad
    \xi_{\bm{k}} (\theta) \ket{0(\theta)} 
    = \Tilde{\xi}_{\bm{k}} \ket{0 (\theta)} = 0\,.
\end{equation}

Then we have the following relation

\begin{equation}
\label{eq:squeezed_vacuum_QFT}
    \ket{0(\theta)} = U_{\text{B}} (\theta) \ket{0}=\exp\qty[-\delta^{(3)} (0) \int \dd^3 {\bm k}\, \ln \cosh \theta_{\bm{k}}] \exp \qty[ \int \dd^3 {\bm k}\, a_{\bm{k}}^\dagger \Tilde{a}_{\bm{k}}^\dagger \tanh \theta_{\bm{k}}] \ket{0}\,, 
\end{equation}
which show that the $(a_{\bm{k}}, \Tilde{a}_{\bm{k}})$ pairs are condensed in the thermal vacuum $\ket{0(\theta)}$. We can construct two types of Fock space $\mathcal{H}(0), \mathcal{H}(\theta)$, which are constructed from vacua $\ket{0}, \ket{0(\theta)}$ and their annihilation and creation operators, respectively. The parameter~$\theta$ is related to the temperature of the system, and the Fock spaces $\mathcal{H} (\theta), \mathcal{H}(\theta')$ with different parameters $\left(\theta \neq \theta'\right)$ are not equivalent in the sense that vectors in one Fock space cannot express the vector in the other. This is due to the divergent factor $\delta^{(3)} (0) = \infty$ in \eqref{eq:squeezed_vacuum_QFT}, which represents the condensation of the infinitely many particles. This is in contrast to the case of quantum mechanics where Hilbert spaces with different $\theta$ are equivalent.\par

The inequivalence of the Fock spaces is a remarkable feature in quantum field theory, and it enables us to describe the various phases in the quantum system. For example, the normal conducting and the superconducting Fock spaces are inequivalent, and in the case of a system with magnetic moments, the different ways of condensation lead to different magnetically ordered phases such as the paramagnetic phase or the ferromagnetic phase \textit{etc}. \par

Let us end this section with the physical interpretation of tilde particles. There is a special relation between tilde and non-tilde systems by requiring the time-independence of the thermal vacuum $\ket{0(\theta)}$. For the free Hamiltonian

\begin{equation}
    H_{\text{tot}} = \int \dd^3 {\bm k}  \qty[\omega_{\bm{k}}^1 a_{\bm{k}}^\dagger a_{\bm{k}} + \omega_{\bm{k}}^2 \Tilde{a}_{\bm{k}}^{\dagger} \Tilde{a}_{\bm{k}}]\,, 
\end{equation}
the time-dependences of the oscillation operators are $a_{\bm{k}}\exp [-i\omega_{\bm{k}}^1 t]$ and $\Tilde{a}_{\bm{k}} \exp [-i\omega_{\bm{k}}^2 t]$ while the time-dependence of the wavefunction pair is given by $\exp [i(\omega_{\bm{k}}^1 +\omega^2_{\bm{k}} )t]$. Therefore, the time-independence of the thermal vacuum requires
\begin{equation}
\label{eq:hole_interpretation_of_tilde_particles}
    \omega_{\bm{k}}^2 = -\omega_{\bm{k}}^1\,. 
\end{equation}
This relation suggests the interpretation of tilde-particles as ``holes" of the non-tilde ones. Consider the system in thermal equilibrium with the heating bath. The absorption of energy from the outside is explained in two ways: the annihilation of the quanta $a$ in the system and the creation of the hole $\Tilde{a}$. 

\section{Bogoliubov transformation and $S$-matrix}
\label{sec:Bogoliubov}
In this section, we construct the $S$-matrix in terms of the Bogoliubov coefficients. We assume the existence of well-defined in and out regions equipped with the annihilation operators $\hat{a}_{\text{in},\alpha}$ and $\hat{a}_{\text{out},\alpha}$ with the commutation relations $[\hat{a}_{\text{in},\alpha},\hat{a}^{\dagger}_{\text{in},\beta}]=[\hat{a}_{\text{out},\alpha},\hat{a}^{\dagger}_{\text{out},\beta}]=\delta_{\alpha,\beta}$ where $\alpha,\beta$ are labels of particles. The Bogoliubov transformation is given by
\begin{align}
    \hat{a}_{\text{out},\alpha}&=\sum_{\beta} ( \hat{a}_{\text{in},\beta} \bar{A}_{\alpha\beta} - \hat{a}^{\dagger}_{\text{in},\beta}\bar{B}_{\alpha\beta})\,, \qquad 
    \hat{a}_{\text{in},\alpha}=\sum_{\beta} (\hat{a}_{\text{out},\beta} A_{\beta\alpha} +\hat{a}^{\dagger}_{\text{out},\beta}\bar{B}_{\beta\alpha})
    \,, 
    \label{eq:Bogdef}
\end{align}
with
\begin{align}
    \sum_{\gamma}(A_{\alpha\gamma}\bar{A}_{\beta\gamma}-B_{\alpha\gamma}\bar{B}_{\beta\gamma})&=\delta_{\alpha,\beta}\,, \qquad
    \sum_{\gamma}(A_{\alpha\gamma}B_{\beta\gamma}-B_{\alpha\gamma}A_{\beta\gamma})=0
    \,, 
    \label{eq:Bog_rel}
\end{align}
By adopting the matrix notation of~\cite{Frolov:1998wf}, they can be represented by
\begin{align}
\hat{\bm{a}}_{\text{out}} = \hat{\bm{a}}_{\text{in}} \bm{A}^+ - \hat{\bm{a}}^\dagger_{\text{in}}\bm{B}^+\,, \quad
\hat{\bm{a}}_{\rm in}=\hat{\bm{a}}_{\rm out}\bm{A}+\hat{\bm{a}}_{\rm out}^{\dagger}\bar{\bm{B}}
\,,
\end{align}
with
\begin{align}
    \bm{A}\bm{A}^+-\bm{B}\bm{B}^+=\bm{I}\,, \qquad
    \bm{A}\bm{B}'-\bm{B}\bm{A}'=\bm{0}
    \,.
\end{align}

Let us follow~\cite{dewitt1975quantum} to find the $S$-matrix. The vacuum states and the normalised multi-particle states are defined by
\begin{align}
\hat{a}_{\text{in},\alpha}\ket{0;\text{in}}&=\hat{a}_{\text{out},\alpha}\ket{0;\text{out}}=0
\label{eq:vacdef}
\,, \\
    \ket{i_{\alpha_1},\cdots,j_{\alpha_n};{\rm in/out}}&:=\frac{1}{\sqrt{i!\cdots j!}}(\hat{a}_{\text{in/out},\alpha_1}^{\dagger})^i\cdots (\hat{a}_{\text{in/out},\alpha_n}^{\dagger})^j\ket{0;{\rm in/out}}
    \,,
\end{align}
where $\alpha_1, \cdots, \alpha_n$ are distinct and the factorials are necessary to accommodate the Bose statistics. In the following, it is convenient to allow repeated labels and write the multi-particle state as
\begin{align}
    \ket{\alpha_1,\cdots,\alpha_n;{\rm in/out}}&:=\hat{a}_{\text{in/out},\alpha_1}^{\dagger}\cdots \hat{a}_{\text{in/out},\alpha_n}^{\dagger}\ket{0;{\rm in/out}}.
\end{align}
In this notation, we have, for example,
\begin{align}
    \ket{i_{\alpha};{\rm in}}=\frac{1}{\sqrt{i!}}\ket{\alpha,\cdots ,\alpha;{\rm in}}\,,
\end{align}
when the labels are repeated $i$ times. The orthogonality and completeness relation read
\begin{align}
\braket{\alpha_1,\cdots,\alpha_n;\text{in/out}}{\alpha_{1'},\cdots,\alpha_{n'};\text{in/out}}&=\delta_{n,n'}\sum_{\substack{\text{all permutation}\\ (\alpha_1,\cdots,\alpha_n)}}\delta_{\alpha_1,\alpha_{1'}}\cdots \delta_{\alpha_n,\alpha_{n'}}
\,, 
\end{align}
and
\begin{align}
\sum_{n=0}^{\infty}\sum_{\alpha_1,\cdots, \alpha_n}\ket{\alpha_1, \cdots ,\alpha_n;{\rm in/out}}\frac{1}{n!}\bra{\alpha_1, \cdots ,\alpha_n;{\rm in/out}}&=\hat{1}
    \,.
    \label{eq:completeness}
\end{align}
The $S$-matrix is defined by the probability amplitudes $\braket{\alpha_1,\cdots,\alpha_{n};\text{out}}{\alpha_{1'},\cdots, \alpha_{n'};\text{in}}$ which can be represented by a unitary operator
\begin{align}
    \hat{S}:=\sum_n \sum_{\alpha_1,\cdots, \alpha_n} \ket{\alpha_1, \cdots ,\alpha_n;{\rm in}}\frac{1}{n!}\bra{\alpha_1, \cdots ,\alpha_n;{\rm out}}
    \,.
\end{align}
The $S$-matrix operator has the properties
\begin{align}
    \braket{\alpha_1,\cdots,\alpha_{n};\text{out}}{\alpha_{1'},\cdots, \alpha_{n'};\text{in}}&=
    \bra{\alpha_1,\cdots,\alpha_{n};\text{in/out}}\hat{S}\ket{\alpha_{1'},\cdots, \alpha_{n'};\text{in/out}}
    \,, \\
    \hat{a}_{\text{in},\alpha}\hat{S}=\hat{S}\hat{a}_{\text{out},\alpha}\,, &\qquad \hat{a}^{\dagger}_{\text{in},\alpha}\hat{S}=\hat{S}\hat{a}^{\dagger}_{\text{out},\alpha}
    \,.
\end{align}

We first study the multi-particle production and annihilation amplitudes:
\begin{align}
    V_{\alpha_1\cdots \alpha_{n}}&:=e^{-iW^0}\braket{\alpha_1,\cdots ,\alpha_n;\text{out}}{0;\text{in}}
    \,, \label{eq:Vdef}\\
    \Lambda_{\alpha_1\cdots \alpha_{n}}&:=e^{-iW^0}\braket{0;\text{out}}{\alpha_1,\cdots, \alpha_n;\text{in}}
    \,, \label{eq:Lamdef}
\end{align}
where $e^{iW^0}$ is the vacuum-to-vacuum transition amplitude
\begin{align}
    e^{iW^0}:=\braket{0;\text{out}}{0;\text{in}}
    \,.
\end{align}
By using the completeness relation \eqref{eq:completeness}, we can solve \eqref{eq:Vdef} and \eqref{eq:Lamdef} for the vacuum states as
\begin{align}
    \ket{0;\text{in}}&=e^{iW^0}\sum_n \sum_{\alpha_1,\cdots, \alpha_n}\frac{1}{n!} V_{\alpha_1\cdots \alpha_{n}} \ket{\alpha_1,\cdots, \alpha_n;\text{out}}
    \,, \label{eq:in_v}
    \\
        \ket{0;\text{out}}&=e^{-i\bar{W}^0}\sum_n \sum_{\alpha_1,\cdots, \alpha_n}\frac{1}{n!} \bar{\Lambda}_{\alpha_1\cdots \alpha_{n}} \ket{\alpha_1,\cdots ,\alpha_n;\text{in}}
        \,.
\end{align}
Substituting these expressions into the definition of the vacuum states \eqref{eq:vacdef} and applying the Bogoliubov transformation \eqref{eq:Bogdef} give
\begin{align}
    0&=e^{iW^0}\sum_n \sum_{\alpha_1,\cdots, \alpha_n,\gamma} \frac{1}{n!}V_{\alpha_1\cdots \alpha_{n}}\left(A_{\gamma\beta}\hat{a}_{\text{out},\gamma}\ket{\alpha_1,\cdots, \alpha_n;\text{out}}
    + \bar{B}_{\gamma\beta}\hat{a}^{\dagger}_{\text{out},\gamma}\ket{\alpha_1,\cdots, \alpha_n;\text{out}}\right)
    \,, \\
    0&=e^{-i\bar{W}^0}\sum_n \sum_{\alpha_1,\cdots, \alpha_n,\gamma}\frac{1}{n!} \bar{\Lambda}_{\alpha_1\cdots \alpha_{n}}\left(\bar{A}_{\beta\gamma}\hat{a}_{\text{in},\gamma}\ket{\alpha_1,\cdots, \alpha_n;\text{in}}
    - \bar{B}_{\beta\gamma}\hat{a}^{\dagger}_{\text{in},\gamma}\ket{\alpha_1,\cdots, \alpha_n;\text{in}}\right)
    \,.
\end{align}
We may solve these equations recursively and find
\begin{align}
    V_{\alpha_1 \cdots \alpha_n}&=
    \begin{cases} 0, &n~\text{odd}\\ \sum_p V_{\alpha_1 \alpha_2}\cdots V_{\alpha_{n-1}\alpha_n}, &n~\text{even}
    \end{cases}
    \label{eq:creationV}
    \\
        \Lambda_{\alpha_1 \cdots \alpha_n}&=\begin{cases} 0, &n~\text{odd}\\ \sum_p \Lambda_{\alpha_1 \alpha_2}\cdots \Lambda_{\alpha_{n-1}\alpha_n}, &n~\text{even}
    \end{cases}
    \label{eq:annihilationLam}
\end{align}
with
\begin{align}
V_{\alpha \beta}&:=-\sum_{\gamma}\bar{B}_{\alpha\gamma}A^{-1}_{\gamma\beta}\,, \quad
    \Lambda_{\alpha\beta}:=\sum_{\gamma}A^{-1}_{\alpha\gamma}B_{\gamma\beta}
    \,,
    \label{eq:VLamdef}
\end{align}
where $\sum_p$ denotes a summation over all distinct permutations of $\alpha_1\cdots \alpha_n$. Note that from \eqref{eq:Bog_rel}, $V_{\alpha\beta}$ and $\Lambda_{\alpha\beta}$ are symmetric in their indices. Eqs.~\eqref{eq:creationV} and \eqref{eq:annihilationLam} show that the general creation/annihilation process can be written as a tensor product of pair creation/annihilation. Then, using the Bogoliubov transformation \eqref{eq:Bogdef} and the relation between the in and out vacua \eqref{eq:in_v}, the $S$-matrix element for one-particle scattering is computed as
\begin{align}
    M_{\alpha\beta}&:=e^{-iW^0}\braket{\alpha;\text{out}}{\beta;\text{in}}
    \nn
    &\,=e^{-iW^0}\bra{\alpha;\text{out}}\hat{a}^{\dagger}_{\text{in},\beta}\ket{0;\text{in}}
    \nn
    &=\sum_n \sum_{\beta_1,\cdot ,\beta_n, \gamma} \frac{1}{n!}V_{\beta_1\cdots \beta_n}\bra{\alpha;\text{out}}\bar{A}_{\gamma\beta}\hat{a}^{\dagger}_{\text{out},\gamma}+B_{\gamma\beta}\hat{a}_{\text{out},\gamma}\ket{\beta_1,\cdots,\beta_n;\text{out}}
    \nn
    &=\bar{A}_{\alpha\beta}+\sum_{\gamma}V_{\alpha\gamma}B_{\gamma\beta}
    \nn
    &= A^{-1}_{\beta\alpha}
    \,.
\end{align}
Here, we have used \eqref{eq:VLamdef} and \eqref{eq:Bog_rel} to obtain the last expression.

The generic process may be understood as a tensor product of pair production, pair annihilation, and one-particle scattering, implying the following expression of the $S$-matrix operator
\begin{align}
    \hat{S}=e^{iW^0} N\exp \left[\frac{1}{2} \hat{\bm{a}}_{\text{out}} \bm{\Lambda} \hat{\bm{a}}'_{\text{out}}
    +\hat{\bm{a}}_{\text{out}}^\dagger (\bm{M}-\bm{I} )\hat{\bm{a}}'_{\text{out}} 
    + \frac{1}{2} \hat{\bm{a}}_{\text{out}}^\dagger \bm{V}\hat{\bm{a}}'^+_{\text{out}}\right] ,
\end{align}
with $\bm{\Lambda}=\bm{A}^{-1}\bm{B}, \bm{V}=-\bar{\bm{B}}\bm{A}^{-1}, \bm{M}=\bm{A}^{-1}{}'$ in the matrix notation. The vacuum-to-vacuum amplitude $e^{iW^0}$ can be determined from
\begin{align}
    1=\braket{0;\text{in}} &= \sum_{n=0}^{\infty}\sum_{\alpha_1,\cdots, \alpha_n}\frac{1}{n!}|\braket{\alpha_1,\cdots ,\alpha_n;\text{out}}{0;\text{in}}|^2 
    \nn
    &= e^{-2\ima W^0} [\det(\bm{I}-\bm{V}\bm{V}^+)]^{-1/2}
    \nn
    &=e^{-2\ima W^0 }[\det(\bm{A}^+\bm{A})]^{1/2}
    \,,
\end{align}
resulting in
\begin{align}
    e^{iW^0}=\theta [\det(\bm{A}^+\bm{A})]^{-1/4}\,, \qquad |\theta|=1
\end{align}
where $\theta$ represents the phase.

\bibliographystyle{utphys.bst}
\bibliography{main}

\end{document}